\documentclass[]{aa}  

\usepackage{graphicx}
\usepackage{txfonts}
\usepackage{CJKutf8} 
\usepackage{textcomp} 

\usepackage[dvipsnames]{xcolor}
\usepackage[colorlinks=true]{hyperref}

\hypersetup{
    colorlinks=true,
    linkcolor=blue,
    filecolor=blue,      
    urlcolor=blue,
    citecolor=teal
}

\newcommand{\zh}[1]{\begin{CJK}{UTF8}{bsmi}#1\end{CJK}}

\newcommand{\pipes}{\textsc{Bagpipes}\xspace}
\newcommand{\prospector}{\textsc{Prospector}\xspace}
\newcommand{\fsps}{\textsc{FSPS}\xspace}
\newcommand{\Mstar}{\ensuremath{\mathrm{M}_\star}\xspace}

\newcommand{\sig}{\ensuremath{\sigma}\xspace}
\newcommand{\eazy}{\textsc{eazy}}

\newcommand{\cosgal}{COS-z5-Q1\xspace}
\newcommand{\gsgal}{GS-z5-Q1\xspace}

\newcommand{\Mgten}{\ensuremath{\mathrm{\Mstar>\rm 10^{10}M}_\odot}\xspace}

\newcommand{\Mglten}{\ensuremath{\mathrm{\Mstar<\rm 10^{10}M}_\odot}\xspace}

\newcommand{\Mgnine}{\ensuremath{\mathrm{\Mstar<\rm 10^{9.0}M}_\odot}\xspace}

\begin{document}

   \title{Double Trouble: Two spectroscopically confirmed low-mass quiescent galaxies at z>5 in overdensities}
   \titlerunning{high-z quiescent galaxies}

   \author{William M. Baker
          \inst{1}\fnmsep\thanks{william.baker@nbi.ku.dk}
          \and
          Kei Ito \inst{2,3}
          \and
          Francesco Valentino \inst{2,3}
          \and 
          Pengpei Zhu (\zh{朱芃佩})\inst{2,3,4}
          \and
          Gianluca Scarpe \inst{2,3}
          \and
          Rashmi Gottumukkala \inst{2,5}
          \and
          Jens Hjorth \inst{1}
          \and
          Laia Barrufet\inst{6}
          \and
          Danial Langeroodi\inst{1}
          }

   \institute{DARK, Niels Bohr Institute, University of Copenhagen, Jagtvej 155A, DK-2200 Copenhagen, Denmark 
        \and
        Cosmic Dawn Center (DAWN), Denmark
        \and 
        DTU Space, Technical University of Denmark, Elektrovej 327, DK-
2800 Kgs. Lyngby, Denmark
        \and
        INAF-Osservatorio Astrofisico di Arcetri, Largo Enrico Fermi 5, I-50125 Firenze, Italy
        \and
        Niels Bohr Institute, University of Copenhagen, Jagtvej 128, DK-2200 Copenhagen North, Denmark
        \and 
         Institute for Astronomy, University of Edinburgh, Royal Observatory, Edinburgh EH9 3HJ, UK
        }


 
  \abstract
   {We present the discovery of two low-mass, high-redshift, quiescent galaxies, GS-z5-Q1 and COS-z5-Q1, using JWST NIRSpec spectroscopy alongside NIRCam and MIRI photometry. Observed at a redshift of z=5.39 and z=5.11 respectively, and with stellar masses of $\rm 10^{9.6}M_\odot$ and $\rm 10^{9.5}M_\odot$, GS-z5-Q1 and COS-z5-Q1 are two of the most distant quiescent galaxies spectroscopically confirmed to-date, and are by far the least massive ($\sim10\times$ lower mass). Full spectrophotometric modelling reveals that COS-z5-Q1 appears to have quenched more than 300Myr prior to observation ($z\sim 7$) and has a formation redshift of around z$\sim$11, whilst GS-z5-Q1 formed and quenched in a single burst around 150Myr prior to observation ($z\sim6$). GS-z5-Q1 is found to lie near the centre of a known high-z overdensity in GOODS-S, as would be expected by galaxy formation models, while COS-z5-Q1 lies towards the outskirts of an overdense region. This highlights the role that environment could play in accelerating galaxy evolutionary processes and could possibly be linked to the galaxies' quiescent nature.
   By modelling their stellar populations, we show that these types of low-mass quiescent galaxies could potentially be descendants of the higher-z "mini-quenched" galaxies. The discovery of these two low-mass $z>5$ quiescent galaxies illuminates a previously undiscovered galaxy population and motivates dedicated follow-up surveys to investigate the overall population.
   }


   \maketitle
%

\section{Introduction}

Quiescent galaxies likely represent an endpoint to the galaxy evolutionary cycle. Shorn of star-forming activity for hundreds of millions of years they provide a compelling test of galaxy evolutionary theory and cosmology \citep[][]{Man2018}.

Prior to JWST, spectroscopically confirmed quiescent galaxies were limited to a few rare examples at $z\sim4$ \citep[e.g.][]{Valentino2020}, but JWST has pushed back the frontier significantly. We now have many $z>4$ spectroscopically confirmed quiescent galaxies \citep[e.g.][]{Carnall2023Nature, Carnall2024, deGraaff2025, Baker2025a, Wu2025, Nanayakkara2025, Weibel2024qgal,Ito2025b, Zhang2025}. The most distant found to-date is at $z=7.3$ in \citet{Weibel2024qgal}. However, the next highest redshift quiescent galaxy found so far is only at $z=4.9$ \citep{deGraaff2025}. 
This means that there is a seeming "redshift desert" between $z=4.9$ and $z=7.3$. However, investigation of quiescent galaxy number densities show that we should still be able to uncover many within this range \citep[e.g.][]{Valentino2023,Alberts2024, Baker2025d}. In addition, photometric candidates have been found \citep[e.g.][]{Alberts2024,Baker2025d, Stevenson2025} but have yet to be followed up with spectroscopy. 

However, one key point is that everything spectroscopically confirmed so far is a massive quiescent galaxy with \Mgten. These are naturally bright and are therefore more readily observed and require less deep spectroscopy. 
Lower-mass high-z quiescent galaxies so far have remained an enigma, even with JWST. Unlike the aforementioned massive quiescent galaxies, the highest redshift spectroscopically confirmed lower mass (\Mglten) quiescent galaxies lie at much lower redshifts \citep[$z<3.5$,][]{Sandles2023,Sato2024,Baker2025a}. Photometric studies uncover many lower-mass (\Mglten) quiescent galaxies \citep{Valentino2023, Alberts2024, Baker2025d}, but due to their intrinsic faintness, spectroscopy for these sources has been lacking.

The most likely of the proposed quenching mechanisms are thought to be different for the massive and low-mass quiescent galaxy sub-populations. 
At lower redshift ($z<2$), lower-mass quiescent galaxies are thought to be quenched primarily by the environment \citep{Dressler1980,Peng2010}, while the culprit for massive quiescent galaxies is generally assumed to be some form of AGN feedback \citep{Croton2006}, either ejective or preventive \citep{Fabian2012, Zinger2020}. Models of quenching can be split into "mass" quenching, primarily affecting massive galaxies \citep[mostly corresponding to AGN feedback in simulations,][]{Vogelsberger2020} and "environmental" quenching \citep{Alberts2022}, typically affecting low-mass galaxies \citep[][]{Peng2010}. But at high-z, the amount of time available in which a galaxy can become quenched in is shorter\footnote{As an example, one cannot quench a $z=7$ galaxy in 1 Gyr due to the age of the Universe at that redshift only giving a absolute maximum of $\sim700$Myr.}. This could impact the ability of the environment to quench low-mass galaxies, as it would require a rapid environmental mechanism \citep[e.g. mergers,][]{Puskas2025}. 

At high redshift, the role of environment is also tightly connected to the growth of protoclusters, the overdense regions that may be the progenitors of the kind of massive clusters observed more locally \citep{Overzier2016,Alberts2022}. Many high-z protoclusters are found \citep[e.g.][]{Laporte2022,Helton2024,Helton2024b,Arribas2024,Fudamoto2025,Witten2025c} with suggestions that the galaxies residing in these protoclusters diverge from field based galaxies, by showing signs of accelerated growth and evolution \citep[][]{Morishita2025,Witten2025b}. This would make these overdense regions the natural place to find quiescent galaxies \citep{Jespersen2025b}, and as expected many high-z spectroscopically confirmed quiescent galaxies show signs of residing in overdense regions \citep{McConachie2022,Tanaka2024,Glazebrook2024,Kakimoto2024, deGraaff2025, Carnall2024, Ito2025, McConachie2025}. This suggests that it is crucial to probe the environment of high-z quiescent galaxies to understand their role within the cosmological context.

An additional source of complexity is the discovery of the so-called "mini-quenched" galaxies \citep{Strait2023, Looser2024, Baker2025b, Covelo-Paz2025}. These high-z low-mass galaxies (\Mgnine), with steep blue UV slopes, weak Balmer breaks, and an absence of strong emission lines, show signs of recent quenching episodes. However, the timescales for these quenching episodes are usually within the last 20-50Myrs \citep{Looser2024,Baker2025b,Covelo-Paz2025}, suggesting only a short period of dormant activity (hence the steep blue UV slopes and weaker Balmer breaks). These are typically thought to result from bursty star-formation and would correspond to periods of possibly temporary quenching before an up tick in star-formation \citep{Dome2024,Gelli2025}.

However, it is an open question as to the role these mini-quenched galaxies play in the galaxy evolutionary cycle. Are they a distinct population or are they a short-lived phase that most galaxies go through?
Could these low-mass mini-quenched galaxies turn into low-mass quiescent galaxies?

In this paper, we confirm the quiescent nature, low-masses, and high redshifts of \gsgal and \cosgal, exploring their stellar populations, environments, and possible progenitors. 
We use a \citet{Chabrier2003} IMF and \citet{PlanckCollaboration2020} cosmology throughout.

\section{Data and methods}

\subsection{The targets}

\begin{figure*}
    \centering
    \includegraphics[width=0.49\linewidth]{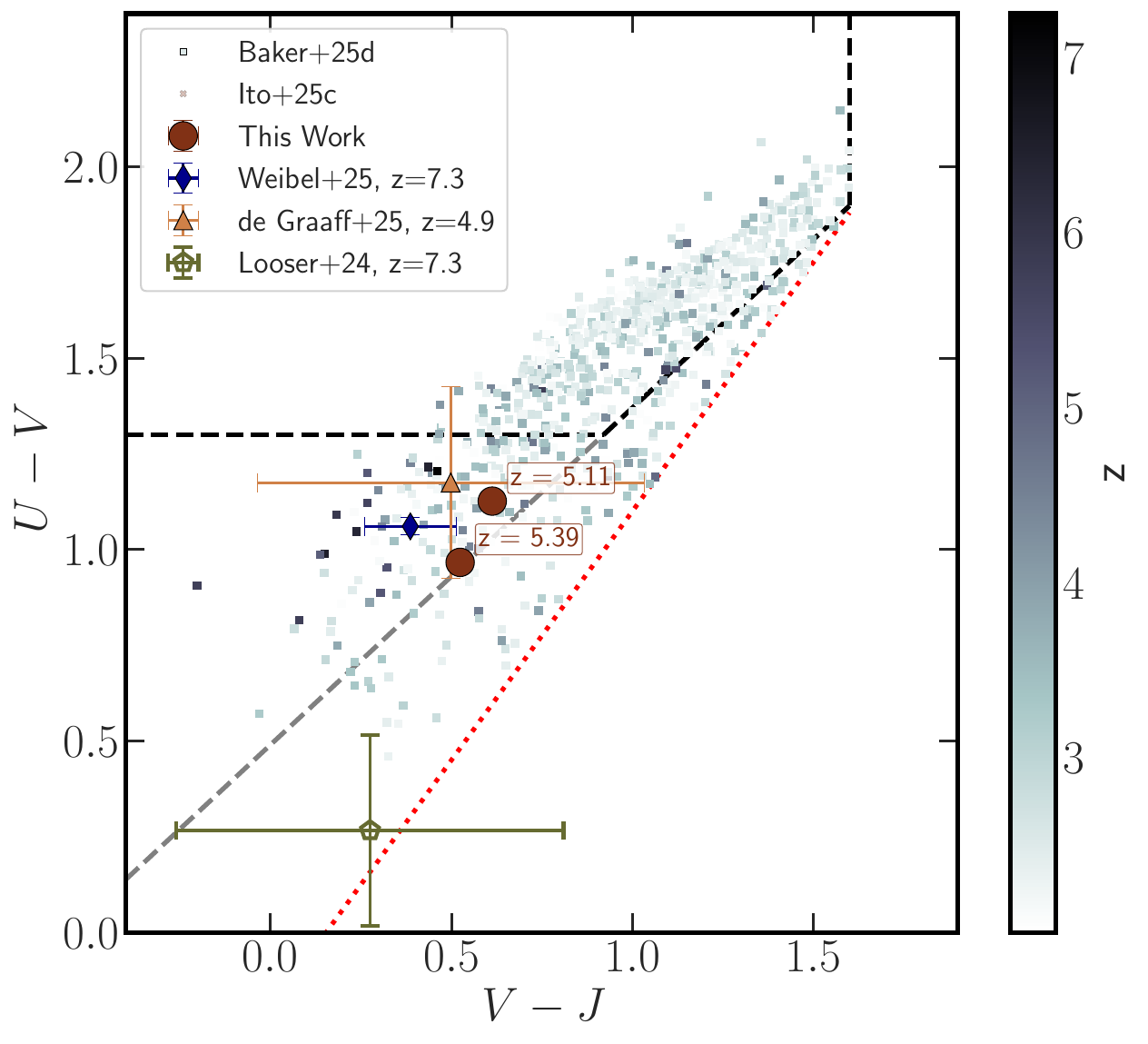}
    \includegraphics[width=0.47\linewidth]{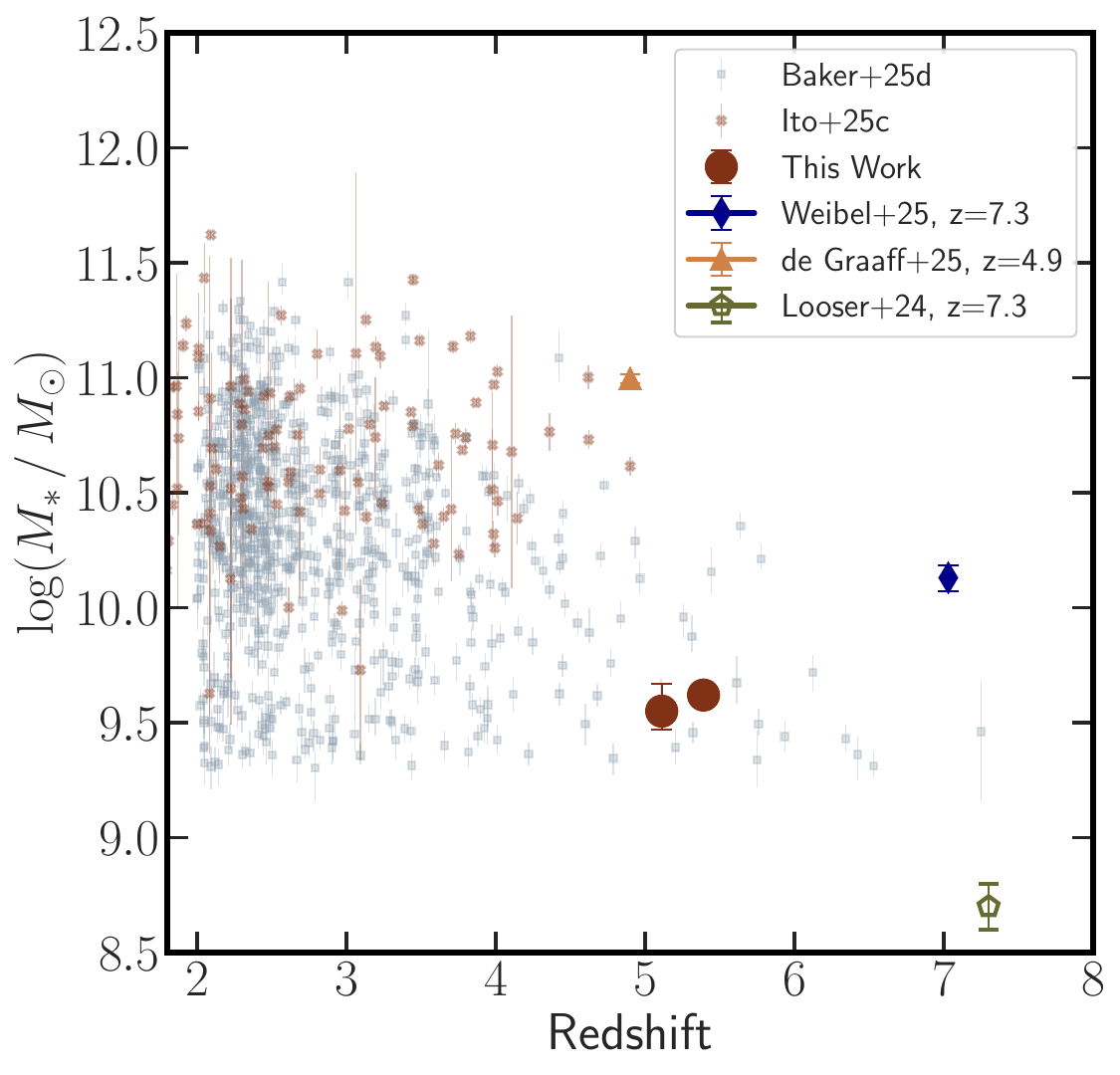}
    \caption{Left: $UVJ$ colour diagram. \gsgal\ and \cosgal\ are indicated with full red circles, along with the sources in \cite{Weibel2024qgal,Looser2024,deGraaff2025} as labelled. The background gray-scaled sources are QG candidates at $z>2$ from \citet{Baker2025d}, where the mass cut has been relaxed ($M_\star>10^{9.3}\,M_\odot$). Right: The same galaxies are shown in the stellar mass versus redshift plane. Quiescent galaxies with JWST medium/high-resolution spectra compiled in \citet{Ito2025b} are also shown in squares. It highlights how low mass \gsgal and \cosgal are compared to the \citet{deGraaff2025} galaxy at $z=4.9$, i.e. almost 1.5\ dex less massive, despite having similar colours and redshifts. }
    \label{fig:uvj}
\end{figure*}

\gsgal and \cosgal were photometrically selected based on the criteria presented in \citet{Baker2025d}. Specifically, their extended selection criteria ($\rm M_*>10^{9.3}M_\odot$) which we briefly summarise as follows. Galaxies were selected from five CANDELS fields \citep{Grogin2011, Koekemoer2011} and template fit using \eazy\ \citep{Brammer2008}. High-z quiescent candidate galaxies (with $z_{\rm phot}>2$) were selected from falling within the extended UVJ based colour selection region defined by \citet{Baker2025a}, which itself was developed based on 18 spectroscopically confirmed massive quiescent galaxies from $2<z<5$. 
This colour selected, reduced, but still large, sample was then modelled with the Bayesian spectral energy distribution (SED) modelling tool \pipes \citep{Carnall2018}. Quiescent galaxies were then selected based on an evolving specific-star-formation rate (sSFR) criterion \citep{Franx2008, Gallazzi2014, Carnall2018, Baker2025a}
\begin{equation}
\label{eq:ssfr_cut}
   \rm  sSFR_{100Myr}\ \leq 0.2/t_{age}
\end{equation}
where $\rm t_{age}$ is the age of the universe at the observed redshift and $\rm sSFR_{100Myr}$ is the specific star-formation rate averaged over the past 100Myr. 

Fig. \ref{fig:uvj} (left) shows the final extended selection sample from \citet{Baker2025d} as the grey-scaled background points on a classic UVJ colour selection diagram. The \citet{Baker2025a} extended UVJ selection criteria is shown by the dotted red line. Also shown is the \citet{Schreiber2015} and \citet{Belli2019} selection criteria as the black dashed lines and grey dashed line respectively. As detailed in \citet{Baker2025d}, they uncover many more low-mass quiescent galaxies with this extended selection followed by their Bayesian-based stellar population modelling approach and sSFR cut (Eq. \ref{eq:ssfr_cut}), than would be selected by the more traditional (low-redshift based) criteria. 

The \citet{Baker2025d} extended photometric catalogue was then cross-matched with the available spectra on the Dawn JWST Archive (DJA\footnote{\url{https://dawn-cph.github.io/dja/}}) and \gsgal and \cosgal were identified. Their positions in the UVJ diagram in Fig. \ref{fig:uvj} show that they fall out of the classic \citet{Schreiber2015} UVJ criteria but fall into the \citet{Belli2019} and \citet{Baker2025a} criteria. \footnote{{It should be noted that these UVJ colours are computed from the \eazy\ best fit model to photometry rather than the spectrum. The UVJ colours obtained by the full spectrophotometric fitting is consistent with that of the \eazy\ colours to around 1$\sigma$. The full spectrophotometric fitting finds that both galaxies are slightly bluer in the the V-J colours. }}
Fig. \ref{fig:uvj} lower panel shows \gsgal and \cosgal in the stellar mass versus redshift plane\footnote{We note that the stellar masses of \gsgal and \cosgal differ by around 0.2dex compared to that measured using photometry by \citet{Baker2025d}, but caution that that is fully consistent with the different star-formation histories and fitting codes used between these two  \citep[e.g.][]{Leja2019}.}. It shows that \gsgal and \cosgal are over an order of magnitude less massive than the \citet{deGraaff2024} $z=4.9$ massive quiescent galaxy despite having similar UVJ colours and residing at a similar redshift. They reside over 0.6dex below even the $z=7.3$ \citet{Weibel2024qgal} galaxy, highlighting just how low-mass these two quiescent galaxies are. Despite this, they remain more massive than the typical high-redshift mini-quenched galaxies by around 0.5-0.7dex \citep{Looser2024, Baker2025b}.

Based on the well-established evolving sSFR criteria shown in  Eq. \ref{eq:ssfr_cut} (alongside other metrics, such as the higher-z UVJ colour selections), both \gsgal and \cosgal are clearly classified as quiescent galaxies. {Due to an almost complete lack of star-formation over the past 100Myrs, they would still be classified as quiescent even after sSFR cuts over shorter timescales (e.g. 50Myrs),} and this point is further enforced by their  strong Balmer breaks in both spectroscopy and photometry, as seen in Fig. \ref{fig:spectra}.

\subsection{Photometry and Spectroscopy}

The photometry and spectroscopy of the targets were both retrieved from the DJA. We adopted the photometry from the most updated v7 version available in PRIMER-COSMOS (v7.4, GO Program ID \#1837, PI: J. Dunlop) and GOODS-S (v7.2). JWST and HST images were reduced with \textsc{Grizli} \citep{Brammer2023} following the procedures detailed in \citet{Valentino2023}. The photometry was extracted in circular apertures with a diameter of $0\farcs5$ and corrected to the total values within an elliptical Kron aperture \cite{Kron1980}. We further applied the zeropoint corrections iteratively computed with \textsc{eazypy} \citep{Brammer2008} and available on the DJA. As noted elsewhere \citep{Ito2025b,Valentino2025}, adopting photometry extracted on PSF-matched images does not introduce any significant colour correction, but can increase the overall normalization by $\approx10$\%. This does not affect the primary conclusions of this work. 
The observed photometry used in the fitting (including the error floor correction) is presented in Table \ref{tab:photometry}.

The PRISM/CLEAR spectra were retrieved from the v4.3 release of DJA\footnote{\url{https://s3.amazonaws.com/msaexp-nirspec/extractions/nirspec_public_v4.3.html}} \citep{Valentino2025, Pollock2025} and reduced with \textsc{msaexp} as described in detail in \cite{deGraaff2025} and \cite{Heintz2025}. \gsgal\ was observed as part of the spectroscopic campaign described in \cite{Barrufet2025} (PID \#2198; PIs: L. Barrufet \& P. Oesch). The low-resolution spectrum ($R\sim100$) was obtained integrating for $2451$s dithered along three microshutters. \cosgal\ was obtained as part of the CANDELS-Area Prism Epoch of Reionization Survey (CAPERS, \citealt{Kokorev2025_capers}; PID \#6368, PI: M. Dickinson). The target benefitted from the deep-tier setup and was observed for $17069$s with the PRISM-CLEAR. The photometry and spectra of both targets are shown in Figure \ref{fig:spectra}.

\begin{table}[]
      \caption[]{Observed photometry of \gsgal and \cosgal.}
         \label{tab:photometry}
         \begin{tabular}{lcc}
            \hline
            \noalign{\smallskip}
                   & \gsgal & \cosgal \\
            \noalign{\smallskip}
            \hline
            \noalign{\smallskip}
 Filter & Flux [nJy] & Flux [nJy] \\
\hline
F090W & 12.6 $\pm$ 2.9 & $-$13 $\pm$ 16 \\
F115W & 25.4 $\pm$ 2.7 & $-$10.2 $\pm$ 7.5 \\
F150W & 29.5 $\pm$ 3.0 & 15.1 $\pm$ 6.5 \\
F200W & 52.2 $\pm$ 5.2 & 21.4 $\pm$ 6.0 \\
F277W & 134 $\pm$ 13 & 78.3 $\pm$ 7.8 \\
F356W & 162 $\pm$ 16 & 89.8 $\pm$ 9.0 \\
F410M & 190 $\pm$ 19 & 93 $\pm$ 11 \\
F444W & 181 $\pm$ 18 & 110 $\pm$ 11 \\
\hline
F560W & 272 $\pm$ 46 & -- \\
F770W & 173 $\pm$ 47 & 27 $\pm$ 140 \\
\hline
F435W & $-$13.9 $\pm$ 4.7 & 12.0 $\pm$ 8.7 \\
F606W & $-$7.4 $\pm$ 3.6 & 11 $\pm$ 12 \\
F814W & 8.4 $\pm$ 3.7 & 31 $\pm$ 12 \\
F775W & $-$4.6 $\pm$ 9.5 & -- \\
 
            \noalign{\smallskip}
            \hline
         \end{tabular}
   \end{table}

\subsection{SED modelling with \prospector}
\label{ss.sed_setup}

We jointly model the spectra and photometry with \textsc{Prospector} \citep{Johnson2021}, which uses the FSPS stellar population synthesis code \citep{Conroy2009,Conroy2010} as its basis. Following the configuration of \citet{Tacchella2022} and similar to that of \citet{Baker2025a}, we adopt a flexible non-parametric continuity SFH \citep{Leja2019} with 12 lookbacktime bins, where we fit for the ratio between the bins. The prior is a student-t distribution with a width of 0.3.
We also use a \citet{Chabrier2003} IMF, and \citet{Calzetti2000} dust law with $A_V$ allowed to vary freely between 0-3 magnitudes. We leave the log stellar metallicity free to vary between -2.0 and 0.19 such that $\log(Z/Z_\odot)=(-2.0,0.19)$.

Redshifts are allowed to freely vary around the spectroscopic value determined with \textsc{msaexp} ($z=5.11$ and $5.39$ for \cosgal\ and \gsgal, respectively, Table \ref{tab:properties}) following a Gaussian with width 0.05.
We also tested leaving redshifts free to vary from $z=1-10$ and found consistent redshifts. We marginalise over the nebular emission lines. Both photometry and NIRSpec spectra are jointly fit, with a 2nd-order polynomial accounting for offsets \citep[for more details on the polynomial fit see][]{Baker2025a}. We adopt a 10\% error floor on the photometry.
Parameter inference is performed via Dynamic Nested Sampling with \textsc{dynesty} \citep{Speagle2020}.

\section{Results}

\subsection{Stellar population properties}

\begin{figure*}
    \centering
    
    \includegraphics[width=0.49\linewidth]{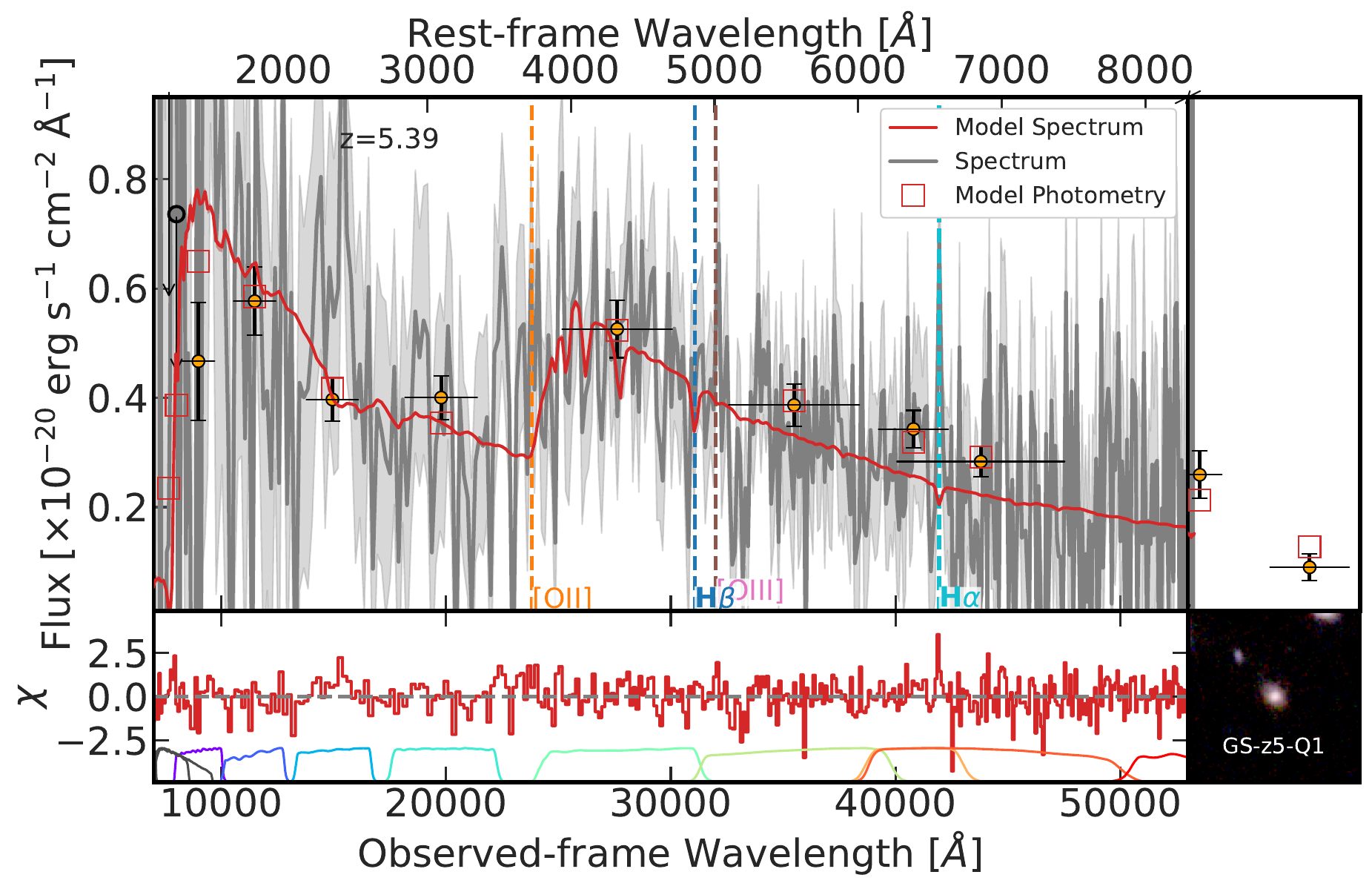}
 \includegraphics[width=0.49\linewidth]{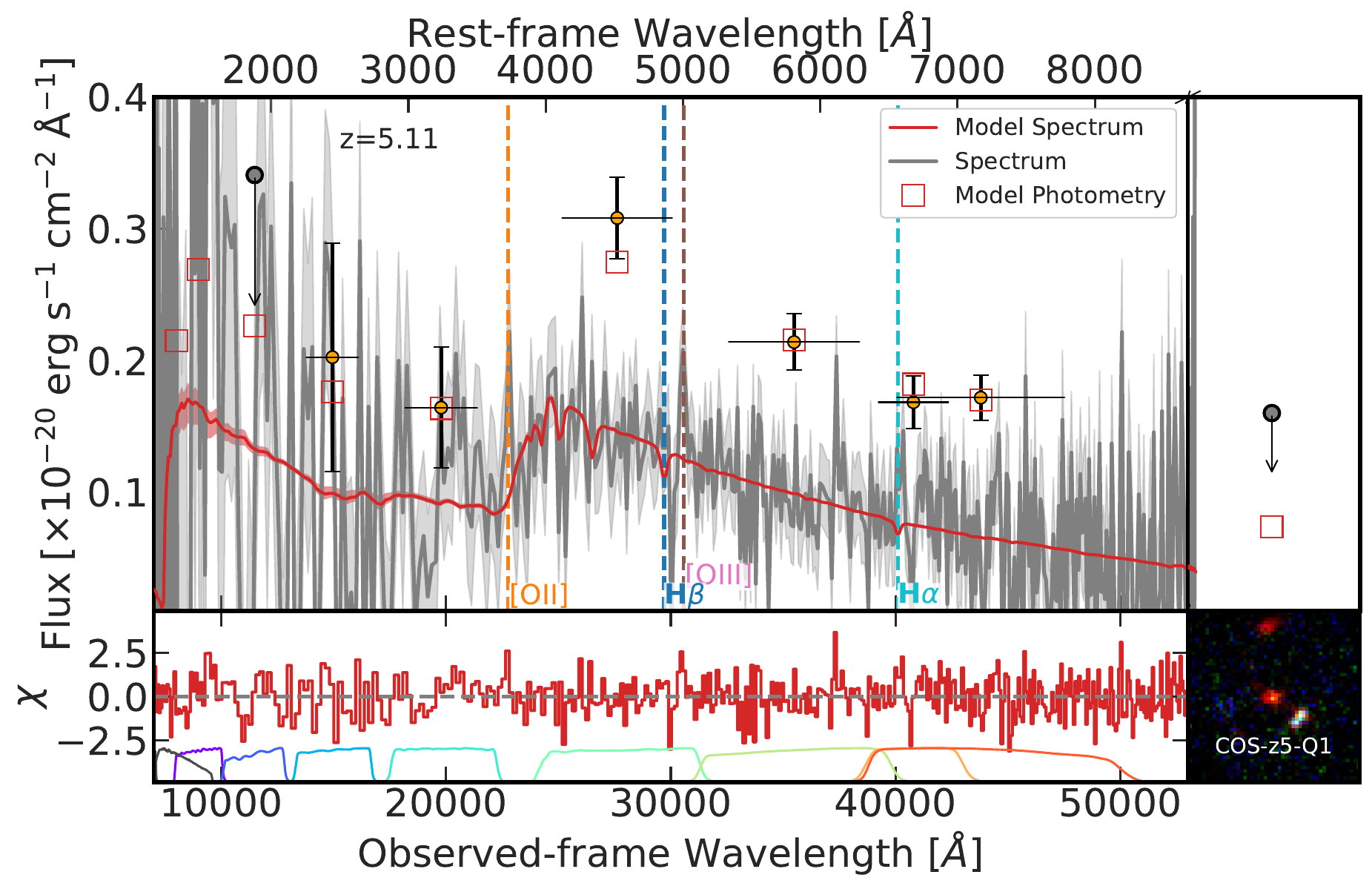}

\includegraphics[width=0.49\linewidth]{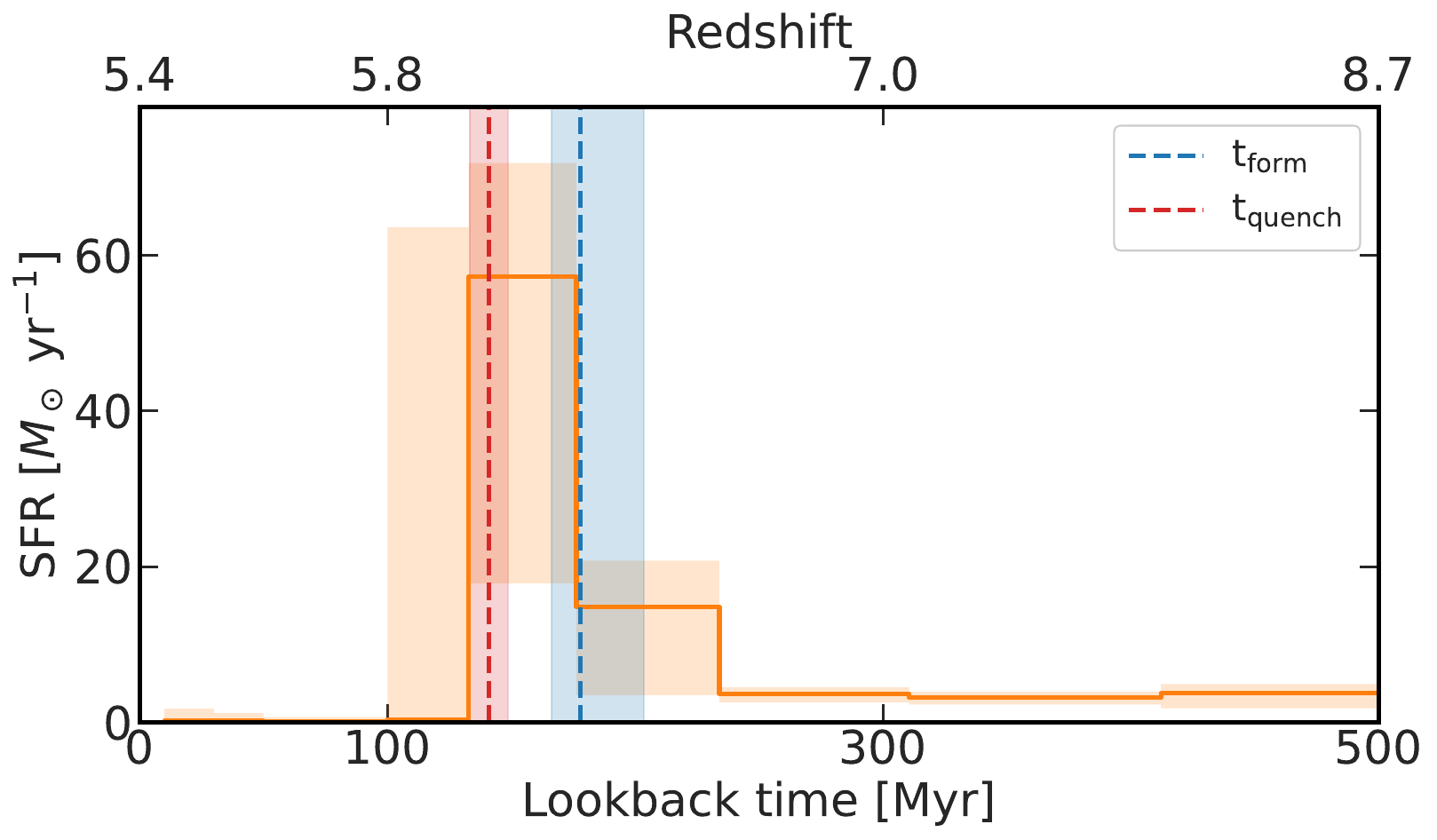}
\includegraphics[width=0.49\linewidth]{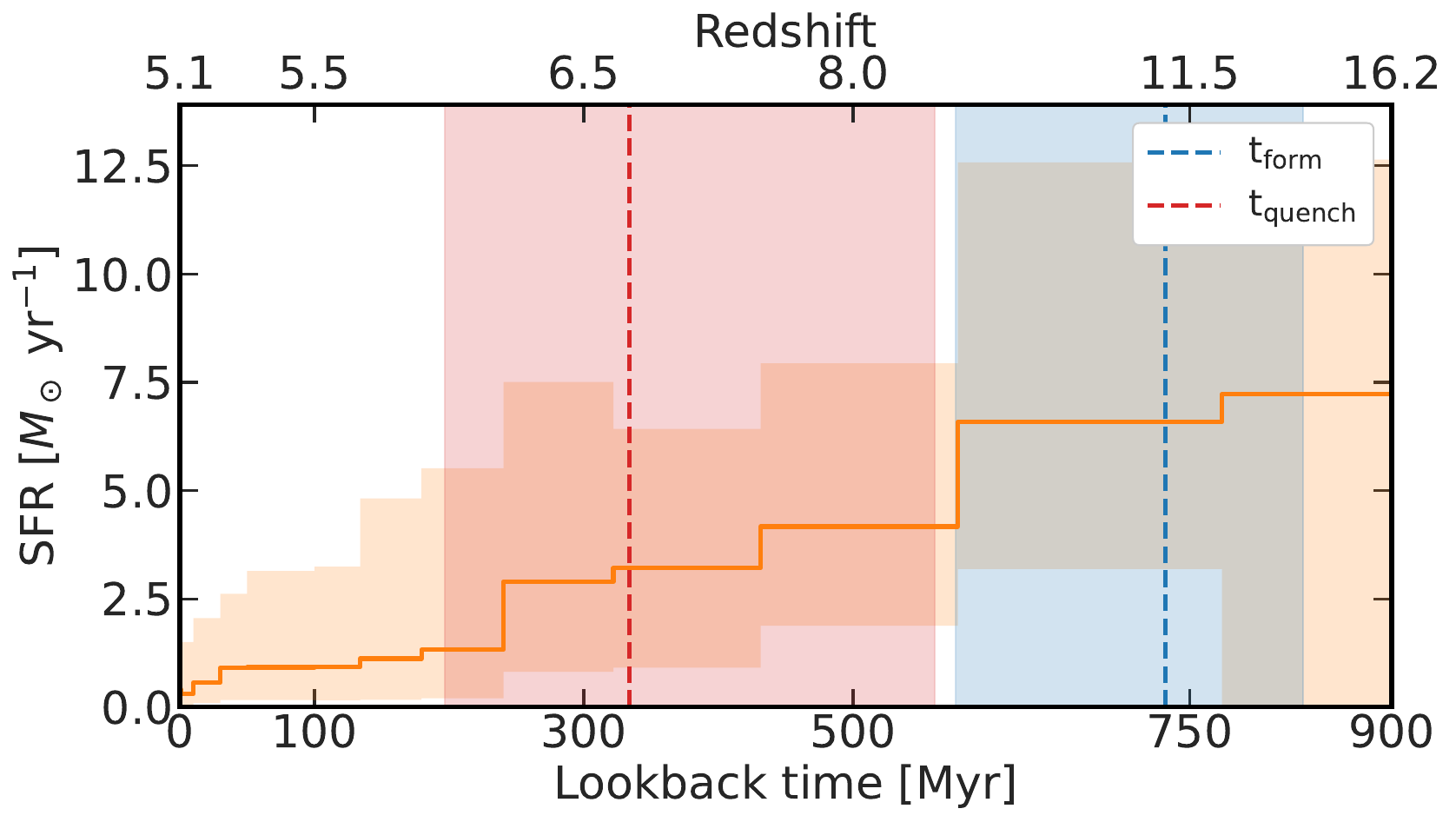}
    
    \caption{Upper: Observed 1D NIRSpec prism spectra (grey), photometry (yellow points) and best fit model spectra and photometry (red) for \gsgal (left) and \cosgal (right). The MIRI points are shown in the right extension to the figure and correspond to F560W (just \gsgal) and F770W (both). Lower middle: $\chi$ values for the spectral fit with the NIRCam filters overplotted (left) and an RGB image of the galaxy (right). Notable emission lines are overplotted. Lower: Star-formation histories for the two galaxies (orange line) showing SFR against lookback time (redshift). The formation time ($\rm t_{50}$, the time 50\% of the mass is formed) and quenching time ($\rm t_{90}$, the time 90\% of the mass is formed) and overplotted as blue and red lines respectively. }
    \label{fig:spectra}
\end{figure*}

Fig. \ref{fig:spectra} upper shows the observed and best-fit model spectra and photometry for \gsgal (left) and \cosgal (right). We see that the spectra and photometry for both sources have strong Balmer breaks again confirming the galaxies' quiescent nature. Neither galaxy shows evidence of strong emission lines, although the position of these lines is overplotted.
We also have additional constraints on the NIR for the two galaxies from the MIRI filter coverage and the NUV from Hubble. This helps rule-out any strong contribution from an AGN component (e.g. a contribution from a hot torus at $\rm \sim1\mu m$). 

The bottom panel of Fig. \ref{fig:spectra} shows the star-formation histories (SFHs) for the two sources. In orange is the 50th percentile of the best-fit SFHs with the errors corresponding to the 16th percentile and 84 percentile. The red and blue lines correspond to the quenching and formation times, which are given by the time taken to form 90\% and 50\% of the total stellar mass respectively.

We can see two contrasting SFHs for the two sources with \cosgal having a much more extended SFH, with significantly greater uncertainty on the formation and quenching times. This is also reflected in the shape of the SFH which might be influenced by the `continuity' SFH prior for this source. Despite this, it would appear to have an early formation and quenching time, consistent with having formed anywhere from $z=9$ to $z=11$ and quenched sometime from $z=6-9$. The SFH for \gsgal appears completely different. The best-fit model indicates a relatively recent short sharp burst of star-formation that accounts for almost the entire SFH. \gsgal has a much younger formation and quenching time of around 150\ Myr prior to observation (e.g. around $z\sim 6$). 

Both SFHs are physically plausible, galaxies with similar $\rm SFR\geq40M_\odot/yr$ to \gsgal have been observed at $z\sim6$. Although far above the normal star-forming galaxy population \citep{Simmonds2025}, it is still well within the region for dusty-star-forming galaxies \citep{Casey2014,Gottumukkala2024} and far below that reached by the most extreme examples \citep[e.g.][]{Zavala2018}. The situation for \cosgal is different as it requires a steady SFR over a longer period, but we do see evidence for galaxies with $\rm SFR\sim5M_\odot$ around $z\sim10$ \citep{Roberts-Borsani2024}. Intriguingly, this opens up the possibility that \cosgal could be a descendent of the UV bright galaxies observed at $z>10$ \citep[e.g.][]{Curtis-Lake2023,Castellano2024,Carniani2025, Naidu2025}.

However, obtaining accurate SFHs and formation and quenching times is complicated with only NIRSpec prism spectra \citep{Baker2025a,Nanayakkara2025, deGraaff2025}. The prism spectra themselves are unable to constrain the stellar metallicity (which is left free to vary within reasonable bounds, see Sec. \ref{ss.sed_setup}), and is itself degenerate with stellar population age. 

Another important caveat to this approach of spectral fitting is that it assumes an entirely in-situ SFH. As highlighted in \citet{Cochrane2025}, mergers are fully expected at high-z due to hierarchical structure formation \citep{White1978}, hence the impact on stellar population properties may be profound. 

However, importantly, we can robustly constrain our galaxy stellar masses to \Mglten with our full-spectrophotometric fitting with \prospector. The reason for this is twofold. Firstly, we are using the "continuity" style SFH prior which is known to increase galaxy stellar masses by 0.1-0.2dex compared with other SFHs \citep{Leja2019, Leja2019stellarmasses}. Secondly, the situation outlined in \citet{Cochrane2025} serves to lead to stellar mass overestimates rather than underestimates. Therefore, should our stellar masses be affected by these two problems, we would expect them to be revised to lower values rather than higher values. This means that the low-mass nature of these two sources should be robust against these SED modelling assumptions.

\begin{table}[]
      \caption[]{Properties of \gsgal and \cosgal.}
         \label{tab:properties}
         \begin{tabular}{lcc}
            \hline
            \noalign{\smallskip}
                   & \gsgal & \cosgal \\
            \noalign{\smallskip}
            \hline
            \noalign{\smallskip}
RA & 53.103040 & 150.15502\\
DEC & -27.83952 & 2.24280\\
$\mathrm{z_{spec}}$ & $5.39^{+0.00}_{-0.00}$ & $5.11^{+0.01}_{-0.01}$ \\ 
$\mathrm{log(M_*/M_\odot}$) & $9.62^{+0.01}_{-0.01}$ & $9.55^{+0.12}_{-0.08}$ \\ 
$\mathrm{SFR_{100}}$ [M$_\odot$/yr] & $0.16^{+0.84}_{-0.16}$ & $0.74^{+1.60}_{-0.61}$ \\ 
$\mathrm{SFR_{10}}$ [M$_\odot$/yr] & $0.01^{+0.12}_{-0.01}$ & $0.39^{+1.53}_{-0.35}$ \\ 
$\mathrm{t_{50}}$ [Gyr] & $0.18^{+0.03}_{-0.01}$ & $0.73^{+0.10}_{-0.16}$ \\ 
$\mathrm{t_{90}}$ [Gyr] & $0.14^{+0.00}_{-0.00}$ & $0.34^{+0.23}_{-0.14}$ \\
$\mathrm{A_v}$ & $0.44^{+0.05}_{-0.04}$ & $0.66^{+0.43}_{-0.33}$ \\ 
$\mathrm{H\alpha\ 10^{-18}[erg/s/cm^2]}$ & $2.56^{+0.30}_{-0.30}$ & $0.50^{+0.08}_{-0.08}$ \\ 
$\mathrm{SFR_{H\alpha}}$ [M$_\odot$/yr] & $0.66^{+0.23}_{-0.23}$ & $0.11^{+0.05}_{-0.05}$ \\ 
$\mathrm{\beta_{UV}}$ & $-0.72^{+0.23}_{-0.23}$ & $-1.63^{+0.32}_{-0.32}$ \\ 
$\mathrm{R_{maj}\ [kpc]}$ & $0.92^{+0.05}_{-0.04}$ & $1.31^{+0.76}_{-0.45}$\\
n & $1.63^{+0.17}_{-0.16}$ & $4.84^{+1.49}_{-1.75}$\\
 
            \noalign{\smallskip}
            \hline
         \end{tabular}
   \end{table}

\subsection{Spectral properties}

In this section, we explore the properties of the NIRSpec spectra of \cosgal and \gsgal. We start by measuring the $\beta_{UV}$ slope. We follow the combined fitting procedure of \citet{Baker2025b} alongside \citet{Saxena2024}. This consists of fitting a powerlaw of the form $\rm F \propto \lambda^\beta_{UV}$ within a window of $1600\AA-2800\AA$. However, we also mask the regions 
$1620\AA-1680 \AA$
and $1860\AA-1980 \AA$ in order to remove potential line contribution \citep{Saxena2024}.
We report the values in Table \ref{tab:properties}.
{We see that both galaxies are consistent with much redder UV slopes, with $\beta_{UV}=-0.72\pm0.23$ for \gsgal, and $\beta_{UV}=-1.63\pm 0.32$ for \cosgal, than are typically seen for galaxies in this redshift range \citep{Roberts-Borsani2024, Heintz2025}. This highlights a lack of recent star-formation when compared to other more recently quenched galaxies, such as in \citet{Looser2024, Baker2025b}.}

We also explore potential emission line contribution. Primarily we are interested in H$\alpha$ as it is such a well known tracer of recent star-formation \citep[on timescales $\sim$10Myr,][]{Kennicutt1998, Kennicutt2012}. This enables us to place a constraint on the recent star-formation activity independently of the SED modelling assumptions. We fit the continuum subtracted prism spectrum with a Gaussian, assuming that all the flux within the H$\alpha$ region is associated with H$\alpha$ (i.e. assuming no blended [NII] contribution). We find a 2.8\sig H$\alpha$ detection.
Converting this into a SFR assuming a \citet{Wilkins2019} conversion factor \citep[although results also remain consistently low with a ][conversion factor]{Kennicutt2012} this gives us an $\rm SFR<1M_\odot\ yr^{-1}$ for both galaxies, as shown in Table \ref{tab:properties}.
This once again highlights the quiescent nature of both \gsgal and \cosgal reiterating that they have not had any recent star-formation.

\subsection{Size and morphology}
The next aspect to explore is the morphologies of these two quiescent galaxies based on the NIRCam imaging. 
We model the surface brightness profiles of the two galaxies with the fitting tool \textsc{PySersic} \citep{Pasha2023}\footnote{\url{https://github.com/pysersic/pysersic}}. The images were retrieved from the DJA and the PSFs from \cite{Genin2025}. 

\begin{figure}
    \includegraphics[width=\columnwidth]{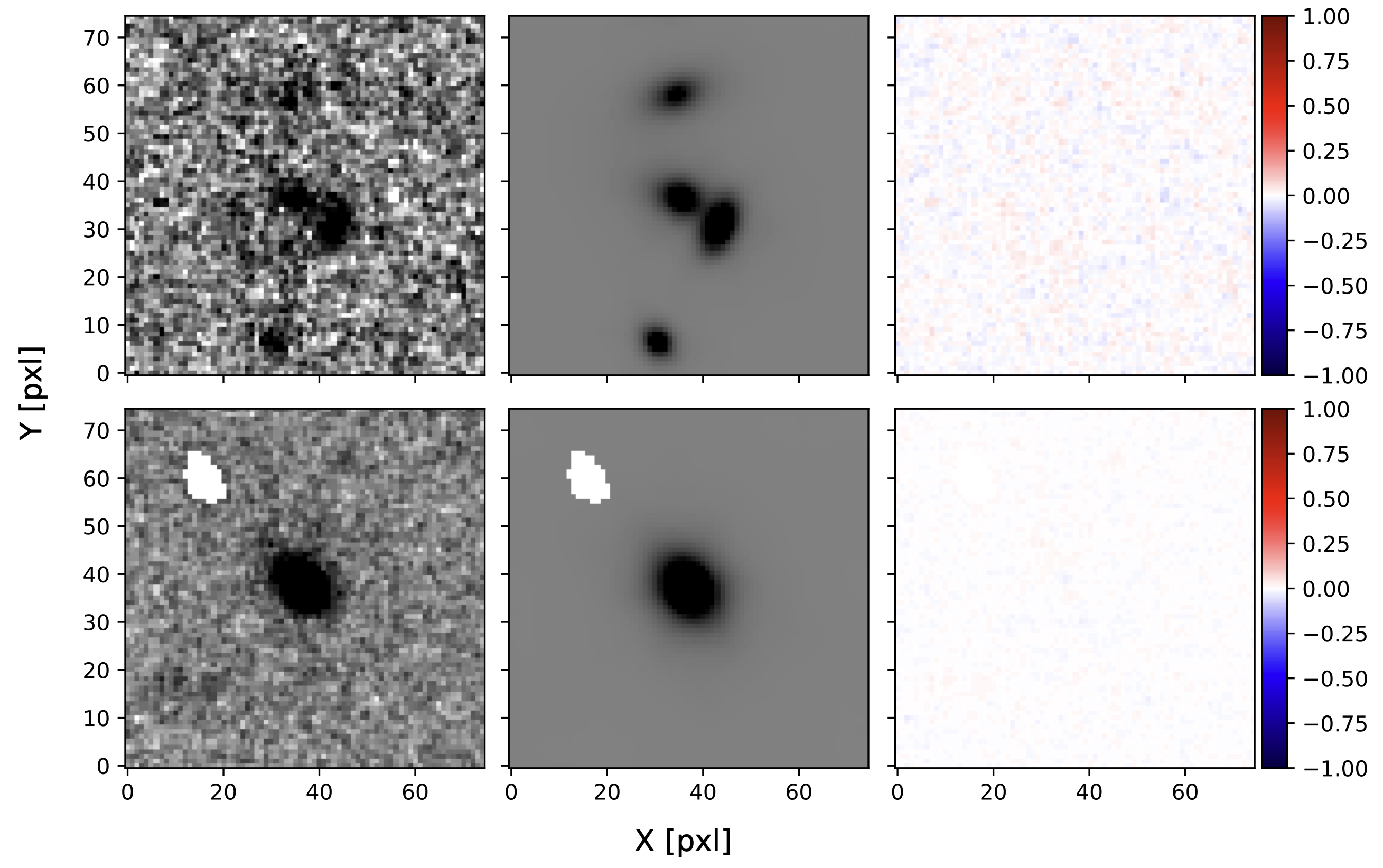}
    \caption{S\'{e}rsic modeling of the surface brightness in NIRCam/F277W of \cosgal\ (top) and \gsgal\ (bottom). In each row, the first panel shows the image in pixel units ($3"$ side, $0\farcs04$ pixel size), the best-fit model, and the residuals.}
    \label{fig:morphology}
\end{figure}

Based on the imaging, \gsgal is both isolated and extended, hence a single \citet{Sersic1968} profile is sufficient to reproduce the observed light profile. The model in each band returns consistent radii and S\'{e}rsic indices, proving that the target is well detected. Figure \ref{fig:morphology} shows the model and the residuals in NIRCam/F277W, corresponding to a rest-frame wavelength of $\approx4500$ \AA. This is the bluest NIRCam wide-band filter available on the red side of the Balmer break. We see no significant residuals.
\gsgal\ is disky ($n=1.63^{+0.17}_{-0.16}$, with an ellipticity $\epsilon = 1-R_{\rm min}/R_{\rm maj}=0.25^{+0.04}_{-0.04}$) and resolved ($R_{\rm maj}=0.92^{+0.05}_{-0.04}$ kpc). Modelling a point source model returns significant structured residuals on extended scales. 

As opposed to \gsgal, the imaging shows that \cosgal is not isolated and has neighbouring sources. These additional sources have different photometric redshifts which are inconsistent with $z=5.11$, hence favouring chance alignment rather than any physical association.
We model \cosgal\ along with the nearby sources to avoid blending and contamination due to imperfect masking (even if minimal). In this case, a simple model with a point source performs only marginally worse than a S\'{e}rsic profile, since the object is compact and faint. The size estimate should thus be taken with a grain of salt and more conservatively considered as an upper limit. For \cosgal, we estimate $R_{\rm maj}=1.31^{+0.76}_{-0.45}$ kpc, a high S\'{e}rsic index $n=4.84^{+1.49}_{-1.75}$ and ellipticity $\epsilon=0.49^{+0.14}_{-0.16}$ in the F277W band.

Neither galaxy shows any sign of central  bulges or cores \citep[e.g.][]{vanDokkum2014,Tacchella2015,Baker2025} and can be fit excellently with single-component models.
Both sources are rather extended compared to the extrapolation of the stellar mass-size relation derived at $z>3$ \citep{Ito2024,Kawinwanichakij2025, Baker2025a}. The effective semi-major axis estimates are also $\sim1.5-5\times$ larger than the sizes reported for the quenched systems at $z\gtrsim5$ discussed in \cite{Weibel2024qgal} and  \cite{deGraaff2025}. However, it is now rather well-established that the mass-size relation tends to flatten at low stellar masses at lower redshifts \citep{Hamadouche2024,Cutler2024}. Coupled with their possible disky structure (especially evident for \gsgal) and benefitting from the deeper imaging available in the field, sources at intermediate masses might represent the connection with the blue star-forming population after quenching, as reported at $3<z<4$ \citep{Sato2024}. 

\section{Discussion}

\subsection{The role of environment}

\begin{figure*}
    \centering
    \includegraphics[width=0.49\linewidth]{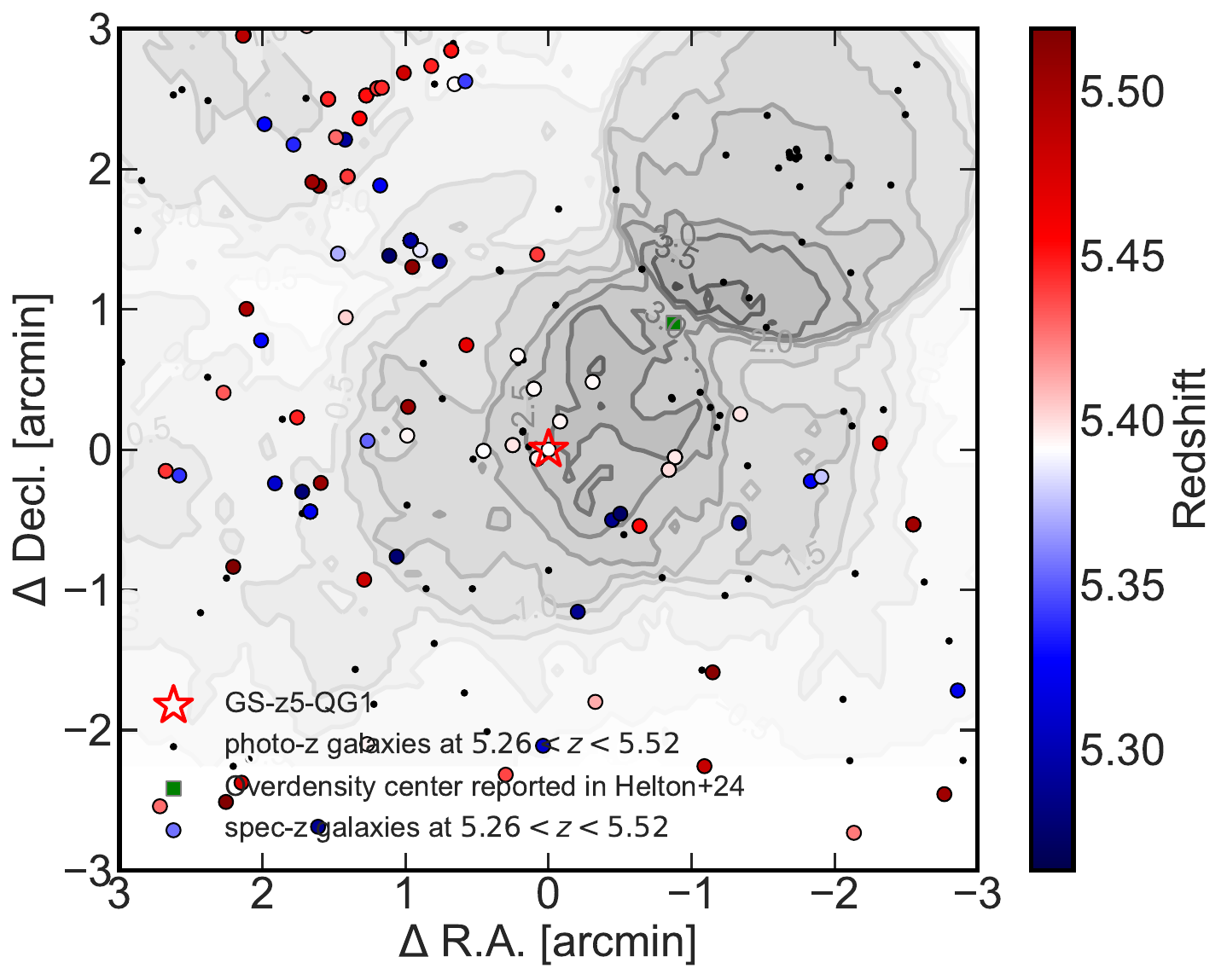}
    \includegraphics[width=0.49\linewidth]{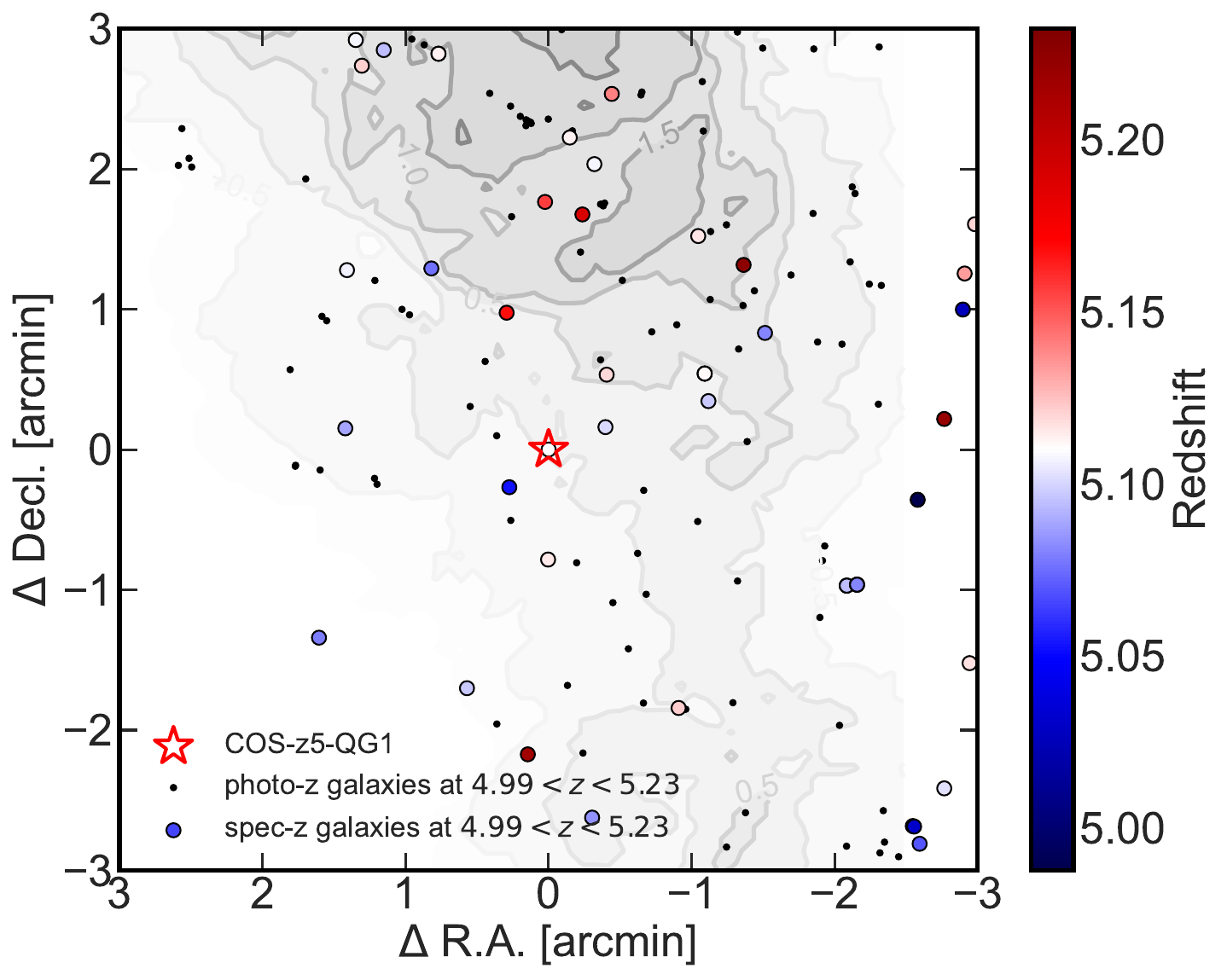}
    \caption{Distribution of galaxies around the \gsgal (left) and \cosgal (right). These QGs are located in the centre of the field and marked as stars. The coloured circles and smaller black circles correspond to galaxies at $5.34<z<5.44$ (for \gsgal) and $5.06<z<5.16$ (for \cosgal) according to their spectroscopic and photometric redshift, respectively. The contours show the overdensity measured from photometric and spectroscopic galaxies. The number in these contours indicates the significance of the overdensity. For \gsgal, the location of the overdensity centre reported in \citet{Helton2024} is shown as a green square.}
    \label{fig:environment}
\end{figure*}

As previously mentioned in the introduction, many quiescent galaxies show signs of lying in overdensities and hence it is important to probe the surrounding environment of \gsgal and \cosgal. 

In this section, we investigate the environment around these two targets. We use the photometric redshift catalogues from the DJA in the GOODS-S and PRIMER-COSMOS field to draw a density map around \gsgal and \cosgal. In order to ensure the quality of the SED fitting and high completeness of the sample in the entire field, we applied a stellar mass cut of $\log{(M_\star/M_\odot)}>8$ and a redshift quality cut of $(z_{\rm phot, 84}-z_{\rm phot, 16})/z_{\rm phot, 50}<1$, where $z_{\rm phot, 16}$, $z_{\rm phot, 50}$, $z_{\rm phot, 84}$ are the 16, 50, 84 percentile of the probability distribution of the photometric redshift. All of these quantities are derived with {\sc eazy-py} \citep[see][for more detail of the configuration]{Valentino2023}. Moreover, we use the spectroscopic redshift of all objects that have a robust (grade=3) estimate in the DJA spectroscopic compilation. To select galaxies that are at the similar redshift to the targets, we considered sources with redshift such that $z_{\rm QG}-\sigma_{NMAD}<z<z_{\rm QG}+\sigma_{NMAD}$, where $z_{\rm QG}$ is the redshift of the target and $\sigma_{NMAD}$ is the normalised median absolute deviation between the photometric and spectroscopic redshifts \citep{Brammer2008}. We also tested varying the threshold to $z_{\rm QG}-0.05<z<z_{\rm QG}+0.05$, and the more traditional $z_{\rm QG}-3\times\sigma_{NMAD}<z<z_{\rm QG}+3\times\sigma_{NMAD}$, finding that the significant overdensities remain regardless of these variation in the threshold. We derived the local surface galaxy density map around \cosgal\ and \gsgal\ by counting galaxies within a $500\, {\rm pkpc}$ aperture at the position. We note that the edge of the field in the survey was masked out during the calculation. The overdensity, i.e., $N-\langle N \rangle$, where $N$ is the number of galaxies within the aperture at the position, is calculated by using the average number $\langle N \rangle$.

Figure \ref{fig:environment} shows the distribution of galaxies around the COS-z5-Q1 and GS-z5-Q1. Interestingly, we find that the GS-z5-Q1 is in a significant overdense structure with $\sim5\sigma$ significance of the peak based on the most constrained redshift window of 0.05. With the wider $\sigma_{NMAD}$ window this decreases to a significance of $\sim3\sigma$ (which is expected since widening the redshift range smooths out the real overdensity.) This dense structure at $z=5.4$ has been previously reported in \citet{Helton2024} where they find 39 members in the wider overdensity. 
Based on the overdensity members, \citet{Helton2024} estimate a halo mass of around   $\rm 12.6<log(M_{halo}/M_{\odot})< 12.8$, but without prism spectroscopy they were unable to confirm \gsgal as a member of this structure. 

As can be seen in the right-hand panel of Fig. \ref{fig:environment}, \cosgal is found to be located at the edge of an overdense structure of galaxies. 
This suggests that both \gsgal and \cosgal may lie within a high-z protocluster.

These results are intriguing, particularly given the low-masses of these two quiescent galaxies compared to other quiescent galaxies at similar redshifts \citep[e.g.][]{deGraaff2025}. 
Locally, low-mass galaxies are commonly thought to be quenched via environmental processes \citep[possibly up to $z\sim2$,][]{Sandles2023} so signs of low-mass quiescent galaxies residing in overdense region opens up the possibility of high-z environmental quenching. 
This could take the form of galaxy harassment \citep{Moore1996}, mergers, or other well-known environmental quenching processes \citep{Alberts2022}.

We note however, just because a low-mass quiescent galaxy lies within an overdense region does not necessitate it was quenched via environmental processes \citep[][]{Alberts2022} and there are many other secular quenching processes that could have led to quiescence\footnote{For example, the possible role of mergers in quenching high-z galaxies is still poorly understood, with simulations predicting a majority of in-situ mass-build-up within massive quiescent galaxies \citep[][]{Baker2025a} and suggesting merging activity is negligible \citep{Chittenden2025}. However, these simulations quickly run into resolution issues so the role of lower mass minor mergers remains unclear.}.
The situation for lower-mass quiescent galaxies is even more poorly understood due to a lack of spectroscopic detections.

Another aspect to consider is that whilst these galaxies are low-mass compared to other detected quiescent galaxies at similar redshifts, when compared to the other galaxies within the overdense region, they are not necessarily particularly `low-mass'. They may even be the most massive galaxy within the potential protocluster, possibly limiting the environmental quenching mechanisms that could have shutdown star-formation. 
This would appear to be less likely for \cosgal due to its position on the edge of the overdensity as opposed to a more central region, However, as previously mentioned (and shown in Fig \ref{fig:environment}) \gsgal lies towards the centre of the overdensity. In addition, in \citet{Helton2024}, where they report and investigate this overdensity that \gsgal resides in, they also report SED modelled stellar masses for the overdensity members that they spectroscopically confirm. They do not report any stellar masses greater than that of \gsgal - their highest reported overdensity member mass is also $M_*\approx 10^{9.6}M_\odot$, i.e. similar to \gsgal. This opens up the possibility that \gsgal is part of a potential protocluster core. 
Further spectroscopy of protocluster members would be required to confirm this scenario.

Even in the case that residing in an overdensity does not mean environmental quenching, it is still intriguing that so many high-z quiescent galaxies lie in such overdense regions \citep[e.g.][]{Glazebrook2017, deGraaff2025, Ito2025, Carnall2024, McConachie2025}. It is thought that protocluster environments can accelerate galaxy evolutionary processes \citep{Morishita2025, Witten2025b} leading to more mature galaxies compared to the field, which may be what we are seeing here for \gsgal and \cosgal. 
Further spectroscopic follow-up would be required to test this.

Another possibility is that this purely relates to their masses. More massive galaxies are expected to form in overdense environments \citep[][]{Jespersen2025b}, whilst (at least more locally) we also see evidence for a mass dependence in quenching \citep{Peng2010, Kawinwanichakij2017}, with more massive galaxies more likely to be quiescent. 

However, another important consideration relating to the number of quiescent galaxies seen in overdensities, is that these regions could be preferentially targeted for follow-up spectroscopy because of their large abundance of interesting galaxies. 
Follow up surveys of high-z quiescent galaxies selected via complete photometric samples will test this scenario.

\subsection{Connection to mini-quenched galaxies}

\begin{figure}
    \centering
    \includegraphics[width=1\linewidth]{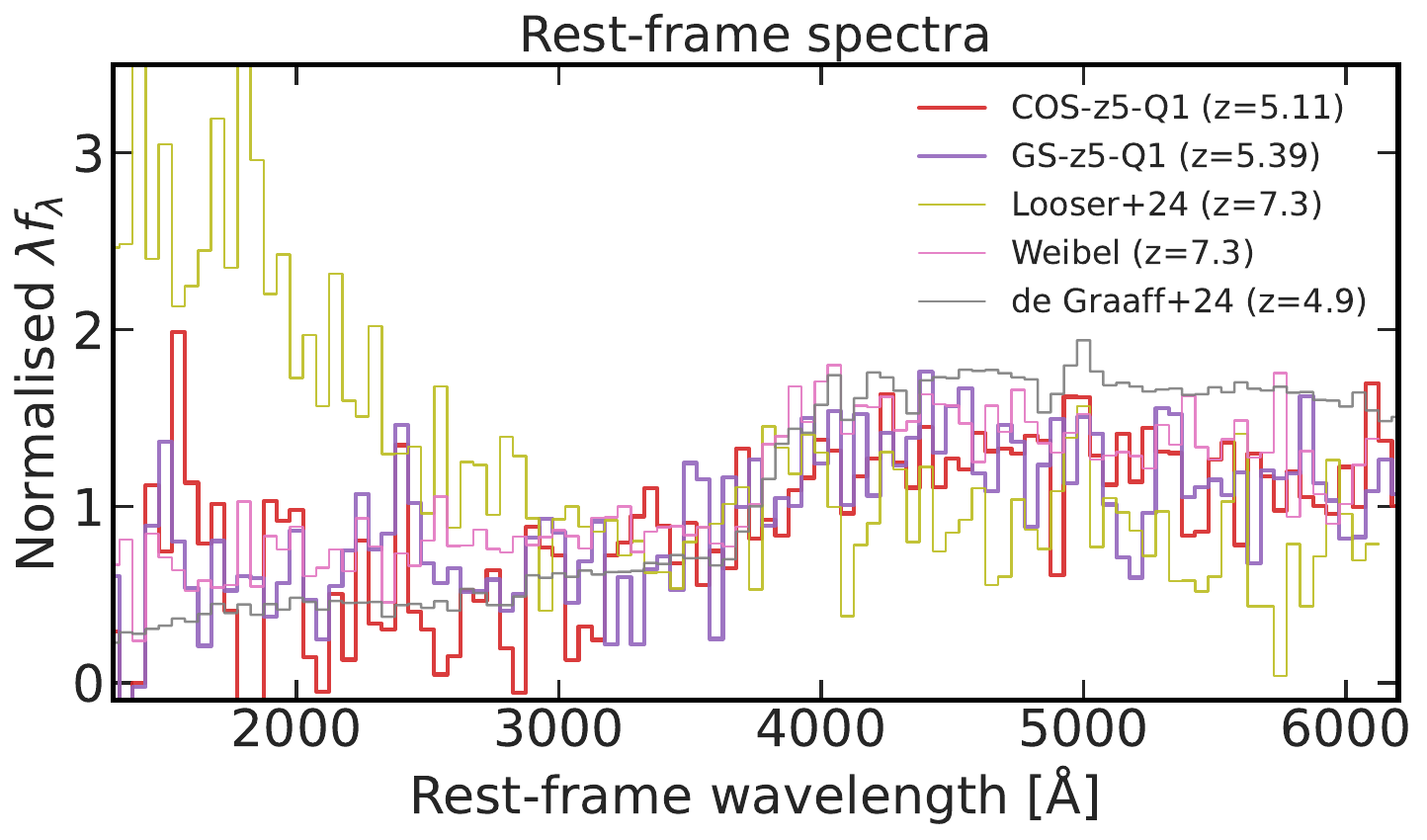}
    \caption{Comparison between \gsgal, \cosgal and three sources from the literature. These are a low-mass "mini-quenched" galaxy from \citet{Looser2024}, a massive quiescent galaxy at $z=4.9$ from \citet{deGraaff2025} and the current highest-redshift massive quiescent galaxy spectrum from \citet{Weibel2024qgal}. The spectra have been rebinned and renormalised to deal better with noise and improve clarity. It shows that \gsgal and \cosgal are consistent with the spectral shapes of the \citet{Weibel2024qgal} and \citet{deGraaff2025} spectra, but are completely different to the classic \citet{Looser2024} "mini-quenched" galaxy spectrum. \gsgal and \cosgal have red UV slopes and more prominent Balmer breaks whereas the \citet{Looser2024} galaxy has a much steeper UV slope and significantly weaker Balmer break. }
    \label{fig:spec_comp}
\end{figure}

One of the most exciting early discoveries with JWST was the finding of the high-z, very low-mass (\Mgnine) recently quenched galaxies, now commonly referred to as "mini-quenched" \citep{Strait2023, Dome2024,Looser2024, Baker2025b, Covelo-Paz2025, Gelli2025}. However, these galaxies present very different spectral quantities compared to what we typically think of as quiescent galaxies. 
 
An example of this is shown clearly in Fig. \ref{fig:spec_comp}, where we present the typical mini-quenched galaxy spectra from \citet{Looser2024} alongside two well-known massive quiescent galaxy spectra from \citet{deGraaff2025} and \citet{Weibel2024qgal}. We also plot the spectra for \gsgal and \cosgal from this work. 

As seen in Fig. \ref{fig:spec_comp}, the differences between the mini-quenched \citet{Looser2024} spectrum and that of the four quiescent galaxies is immediately apparent. The \citet{Looser2024} galaxy has a significantly bluer UV slope with a value of $\beta=-2.09$, alongside a weaker Balmer break compared to all four quiescent galaxies. Other mini-quenched galaxies have been shown to have even bluer $\beta$ slopes than the \citet{Looser2024} galaxy, with the highest redshift example reaching $\beta=-2.8$ \citep{Baker2025b}. This blue UV slope is seen in the UVJ diagram in Fig. \ref{fig:uvj} (upper panel) where we see the extremely blue U-V colour of the \citet{Looser2024} galaxy compared to the other 4 quiescent galaxy sources.

This can easily be understood by the more recent star-formation and quenching times of the mini-quenched galaxies compared to the much longer quenched period of the quiescent galaxies $>100\rm Myr$. The UV slope traces star formation on a 50-100Myr timescale \citep[among other properties, e.g.][]{Bruzual2003,Wilkins2011} hence is much steeper in galaxies with recent SF within the last 100Myrs. The opposite is true for the Balmer break, it can be washed out by recent star-formation, in a process known as "outshining" \citep{Sawicki1998}\footnote{Therefore it is difficult to determine if the mini-quenched galaxies host any older stellar populations \citep{Baker2025b, Witten2025}.}.

An open question remains as to what happens to these mini-quenched galaxies after observation. Their existence is commonly explained as a product of bursty star-formation \citep[e.g.][]{Looser2025, Endsley2024,Endsley2025, Langeroodi2024, Dome2024} whereby they represent the downturn stage of the cycle \citep{Trussler2025, Gelli2025}. This implies that they remain mini-quenched for a short period before an upturn in star-formation occurs and they reignite \citep{Witten2025}.
However, the finding in this work of two lower-mass massive quiescent galaxies opens up the intriguing possibility that the mini-quenched galaxies observed at higher-z are close progenitors to the low-mass quiescent galaxies.
\begin{figure*}
    \centering
    \includegraphics[width=0.9\linewidth]{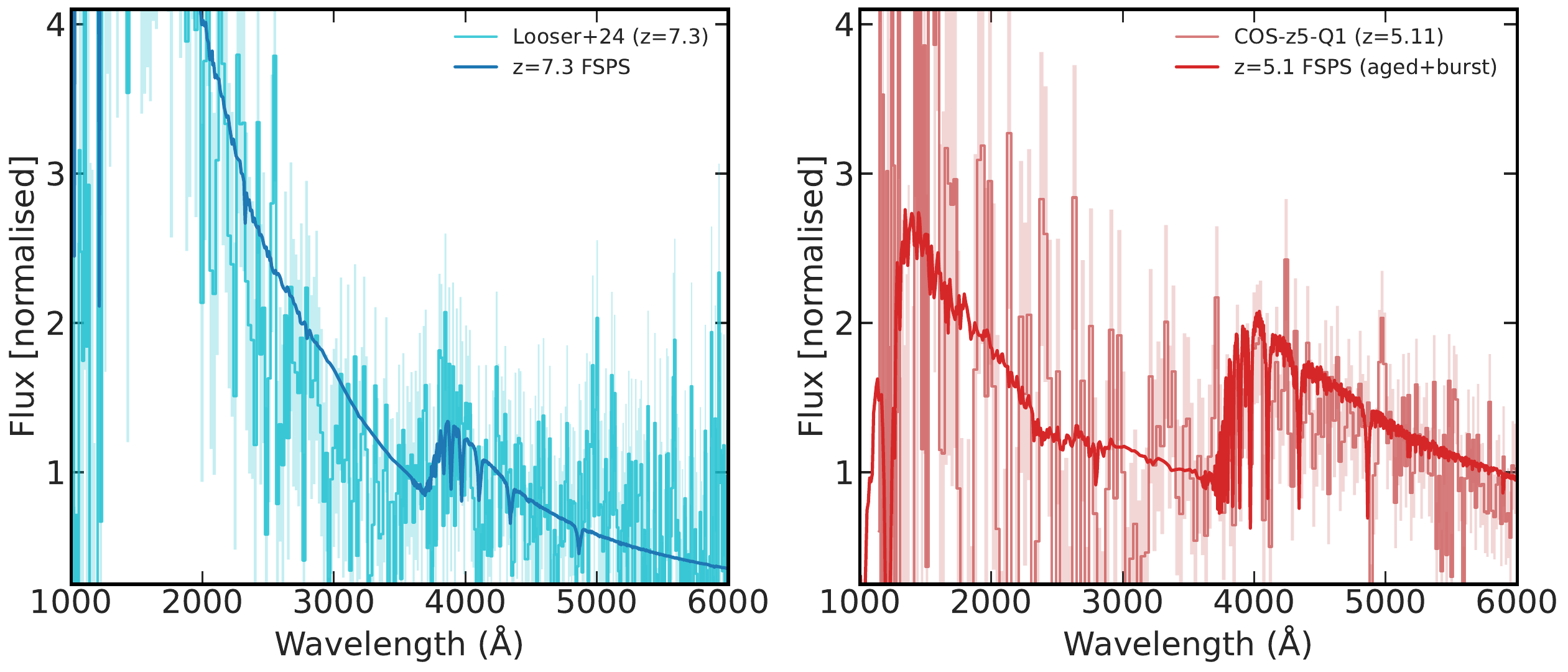}
    \caption{Left: the $z=7.3$ \citet{Looser2024} mini-quenched galaxy spectrum with an appropriate same redshift \fsps model overplotted (red). Right: \cosgal observed spectrum with the same \fsps model, only with the addition of a burst and with stellar population aged from $z=7.3$ to $z=5.1$. It shows that it is possible to the evolution of a mini-quenched galaxy into a low-mass quiescent galaxy with minimal assumptions. }
    \label{fig:fsps_miniquenched}
\end{figure*}

With this in mind we now explore whether the \citet{Looser2024} galaxy could turn into a lower-mass quiescent galaxy such as \cosgal or \gsgal. We cannot just let the stellar populations in the \citet{Looser2024} galaxy age as that galaxy has a mass of $10^{8.7}M_\odot$ and our lowest mass quiescent galaxy has $10^{9.5}M_\odot$. This implies we require a 0.8dex growth in the stellar mass, hence we require some amount of star-formation rejuvenation\footnote{However, it should be noted that were the \citet{Looser2024} mini-quenched galaxy to remain completely dormant and turn into a $10^{8.7}M_\odot$ quiescent galaxy it would develop a much stronger Balmer break and redder UV slope (stronger than \cosgal, which is consistent with \cosgal having a more recent SFH). The difficulty is that it would be so faint as to be undetectable within any reasonable integration time with the NIRSpec instrument. This highlights one reason why we have not found $z\sim5$, $M_*\approx 10^{8.5}M_\odot$ quiescent galaxies. Therefore, we will not probe this purely ageing scenario, rather the possibility that the \citet{Looser2024} galaxy could turn into a quiescent galaxy with the mass of \cosgal. It is worth noting that this could open up the possibility of finding $\rm M_*\approx 10^{9.5}M_\odot$ mini-quenched galaxies at high-z.}.

The first thing to look at is the star-formation histories of \cosgal and \gsgal. \gsgal appears more consistent with being recently formed and quenched, and whilst this does not rule it out as a descendent (any previous SF before the burst could be thoroughly erased), it does suggest the \gsgal progenitor is likely to have much higher star-formation rates \citep[e.g. some form of dusty star-forming galaxy,][]{Casey2014}. Therefore, we focus on \cosgal which has a quenching time of around 300Myr. This quenching time is consistent with observations of the $z=7.3$ massive quiescent galaxy in \citet{Weibel2024qgal} and indirectly with dust removal timescales at $z\sim5$ \citep{Lesniewska2025}. This highlights that this quenching timescale is plausible for quiescent galaxies at this redshift. 

In order to model the galaxies we use \fsps \citep{Conroy2009, Conroy2010} in order to create two $z=7.3$ stellar populations in order to mimic the spectrum of the \citet{Looser2024} mini-quenched galaxy. The age of the stellar populations correspond to the formation and quenching time of the galaxy (as inferred by their \prospector fit) respectively. This \fsps model is shown alongside the normalised \citet{Looser2024} spectrum in the left hand panel of Fig. \ref{fig:fsps_miniquenched}. 

We then produce the same model at z=5.1, where the stellar populations are aged accordingly. In order to match the stellar mass of \cosgal we introduce an extra stellar population with an age of 300 Myr (corresponding to the \cosgal quenching time) such that the overall stellar population model has the same mass as \cosgal. The resulting \fsps model is shown in the right-hand panel of Fig. \ref{fig:fsps_miniquenched} alongside the spectrum of \cosgal. We see that this model provides a qualitatively good fit to the normalised \cosgal spectrum, with a significantly redder UV slope and a much stronger Balmer break compared to the original mini-quenched model shown in the left hand panel.

The point of this exercise is not to confirm its progenitor nature, rather show that under simple modelling assumptions it is possible to produce a high-z quiescent galaxy (such that we observe in this work) from a lower mass mini-quenched galaxy that has been observed at a higher redshift.

A full statistical analysis of these two populations (high-z mini-quenched galaxies and low-mass quiescent galaxies) will require much larger spectroscopic samples and motivates dedicated follow-up programmes. This is likely to be addressed on the mini-quenching side by programmes such as OASIS (PID:\#5997, PI: T. Looser, Looser+, in prep) and Sleeping Beauties (PID:\#7511, PI: A. Covelo-paz \& P. Oesch), but requires the addition of dedicated programmes to uncover low-mass high-z quiescent galaxy population.

\section{Conclusions}

In this work we have presented \gsgal and \cosgal, two low-mass high-z ($z>5$) spectroscopically confirmed quiescent galaxies. We have modelled their stellar population properties, morphologies, and explored their environments.
Our conclusions are as follows:

\begin{itemize}
    \item We report the discovery of two of the highest redshift quiescent galaxies. With redshifts of $z=5.39$ and $z=5.11$ \gsgal and \cosgal are two of the most distant quiescent galaxies for which we have NIRSpec spectroscopy.\\

    \item \gsgal and \cosgal are by far the lowest-mass high-z galaxies probed to-date with stellar masses of $10^{9.6}M_\odot$ and $10^{9.5}M_\odot$ respectively. This is almost an order of magnitude less than other comparable spectroscopically confirmed quiescent galaxies at similar redshifts and 0.6\ dex lower than the highest-redshift spectroscopically confirmed quiescent galaxy known.\\

    \item The star-formation histories for \gsgal and \cosgal are very different with \cosgal consistent with an extended SFH, with a formation redshift of $z\sim10$ and a quenching redshift of $z\sim7$, whilst \gsgal appears to have formed and quenched in a single burst approximately 150Myr prior to observation at $z\sim6$. This suggests two different progenitors and quenching mechanisms for these galaxies.\\
    
    \item Exploring their environments, we confirm \gsgal lies at the heart of a known overdensity at $z=5.39$ in GOODS-S. We show that \gsgal lies at the centre of this overdense possible protocluster region, suggesting environment may be contributing to its accelerated evolution. In addition \cosgal also lies at the edge of an overdense region. This opens up the possibility of environmental quenching playing a role at high-z, or at least accelerating galaxy evolution compared to the field.\\

     \item We show that the higher-z, lower mass "mini-quenched" galaxies are possible progenitors to the kind of low-mass quiescent galaxies we have uncovered with minimal modelling assumptions. This suggests that the mini-quenched galaxies could be close to a becoming quiescent for an extended period of time.
    
\end{itemize}

The discovery of these two low-mass $z>5$ quiescent galaxies illuminates a previously undiscovered galaxy population. Follow-up analysis and surveys of these systems and other candidates is crucial to solve the mystery of low-mass galaxy quenching and the possible role of the environment within the process.

\section*{Data Availability}
 The raw JWST data is available via the DAWN JWST Archive (\href{https://dawn-cph.github.io/dja/index.html}{DJA}). 

\begin{acknowledgements}

WMB would like to acknowledge support from DARK via the DARK Fellowship. This work was supported by a research grant (VIL54489) from VILLUM FONDEN.
Some of the data products presented herein were retrieved from the Dawn JWST Archive (DJA). DJA is an initiative of the Cosmic Dawn Center, which is funded by the Danish National Research Foundation under grant DNRF140. KI, FV, and PZ acknowledge support from the Independent Research Fund Denmark (DFF) under grant 3120-00043B. This study was supported by JSPS KAKENHI Grant Number JP23K13141.

We acknowledge use of \textsc{astropy} \citep{AstropyCollaboration2013}, \textsc{fsps} \citep{Conroy2009, Conroy2010}, \eazy\ \citep{Brammer2008}, \textsc{grizli} \citep{Brammer2023}, \textsc{pysersic} \citep{Pasha2023}, \textsc{numpy} \citep{harris2020array}, \textsc{dynesty} \citep{Speagle2020}, 
\textsc{matplotlib} \citep{Hunter:2007}, and \textsc{topcat} \citep{Taylor2005}.

This work is based [in part] on observations made with the NASA/ESA/CSA James Webb Space Telescope. The data were obtained from the Mikulski Archive for Space Telescopes at the Space Telescope Science Institute, which is operated by the Association of Universities for Research in Astronomy, Inc., under NASA contract NAS 5-03127 for JWST. These observations are associated with program \#2198 \citep{Barrufet2025}, \#6368 \citep{Kokorev2025_capers}, \#1180, \#1210, and \#1286 \citep{Eisenstein2023, Rieke2023}, and \#1837 \citep{Donnan2024}.

\end{acknowledgements}

\bibliographystyle{aa}
\bibliography{refs}

\begin{thebibliography}{119}
\expandafter\ifx\csname natexlab\endcsname\relax\def\natexlab#1{#1}\fi

\bibitem[{{Alberts} \& {Noble}(2022)}]{Alberts2022}
{Alberts}, S. \& {Noble}, A. 2022, Universe, 8, 554

\bibitem[{{Alberts} {et~al.}(2024){Alberts}, {Williams}, {Helton}, {Suess}, {Ji}, {Shivaei}, {Lyu}, {Rieke}, {Baker}, {Bonaventura}, {Bunker}, {Carniani}, {Charlot}, {Curtis-Lake}, {D'Eugenio}, {Eisenstein}, {de Graaff}, {Hainline}, {Hausen}, {Johnson}, {Maiolino}, {Parlanti}, {Rieke}, {Robertson}, {Sun}, {Tacchella}, {Willmer}, \& {Willott}}]{Alberts2024}
{Alberts}, S., {Williams}, C.~C., {Helton}, J.~M., {et~al.} 2024, \apj, 975, 85

\bibitem[{{Arribas} {et~al.}(2024){Arribas}, {Perna}, {Rodr{\'\i}guez Del Pino}, {Lamperti}, {D'Eugenio}, {P{\'e}rez-Gonz{\'a}lez}, {Jones}, {Crespo G{\'o}mez}, {Curti}, {Lim}, {{\'A}lvarez-M{\'a}rquez}, {Bunker}, {Carniani}, {Charlot}, {Jakobsen}, {Maiolino}, {{\"U}bler}, {Willott}, {B{\"o}ker}, {Chevallard}, {Circosta}, {Cresci}, {Kumari}, {Parlanti}, {Scholtz}, {Venturi}, \& {Witstok}}]{Arribas2024}
{Arribas}, S., {Perna}, M., {Rodr{\'\i}guez Del Pino}, B., {et~al.} 2024, \aap, 688, A146

\bibitem[{{Astropy Collaboration} {et~al.}(2013){Astropy Collaboration}, {Robitaille}, {Tollerud}, {Greenfield}, {Droettboom}, {Bray}, {Aldcroft}, {Davis}, {Ginsburg}, {Price-Whelan}, {Kerzendorf}, {Conley}, {Crighton}, {Barbary}, {Muna}, {Ferguson}, {Grollier}, {Parikh}, {Nair}, {Unther}, {Deil}, {Woillez}, {Conseil}, {Kramer}, {Turner}, {Singer}, {Fox}, {Weaver}, {Zabalza}, {Edwards}, {Azalee Bostroem}, {Burke}, {Casey}, {Crawford}, {Dencheva}, {Ely}, {Jenness}, {Labrie}, {Lim}, {Pierfederici}, {Pontzen}, {Ptak}, {Refsdal}, {Servillat}, \& {Streicher}}]{AstropyCollaboration2013}
{Astropy Collaboration}, {Robitaille}, T.~P., {Tollerud}, E.~J., {et~al.} 2013, \aap, 558, A33

\bibitem[{{Baker} {et~al.}(2025{\natexlab{a}}){Baker}, {D'Eugenio}, {Maiolino}, {Bunker}, {Simmonds}, {Tacchella}, {Witstok}, {Arribas}, {Carniani}, {Charlot}, {Chevallard}, {Curti}, {Curtis-Lake}, {Jones}, {Kumari}, {Rinaldi}, {Robertson}, {Williams}, {Willott}, \& {Zhu}}]{Baker2025b}
{Baker}, W.~M., {D'Eugenio}, F., {Maiolino}, R., {et~al.} 2025{\natexlab{a}}, \aap, 697, A90

\bibitem[{{Baker} {et~al.}(2025{\natexlab{b}}){Baker}, {Lim}, {D'Eugenio}, {Maiolino}, {Ji}, {Arribas}, {Bunker}, {Carniani}, {Charlot}, {de Graaff}, {Hainline}, {Looser}, {Lyu}, {Rinaldi}, {Robertson}, {Schaller}, {Schaye}, {Scholtz}, {{\"U}bler}, {Williams}, {Willmer}, {Willott}, \& {Zhu}}]{Baker2025a}
{Baker}, W.~M., {Lim}, S., {D'Eugenio}, F., {et~al.} 2025{\natexlab{b}}, \mnras, 539, 557

\bibitem[{{Baker} {et~al.}(2025{\natexlab{c}}){Baker}, {Tacchella}, {Johnson}, {Nelson}, {Suess}, {D'Eugenio}, {Curti}, {de Graaff}, {Ji}, {Maiolino}, {Robertson}, {Scholtz}, {Alberts}, {Arribas}, {Boyett}, {Bunker}, {Carniani}, {Charlot}, {Chen}, {Chevallard}, {Curtis-Lake}, {Danhaive}, {DeCoursey}, {Egami}, {Eisenstein}, {Endsley}, {Hausen}, {Helton}, {Kumari}, {Looser}, {Maseda}, {Pusk{\'a}s}, {Rieke}, {Sandles}, {Sun}, {{\"U}bler}, {Williams}, {Willmer}, \& {Witstok}}]{Baker2025}
{Baker}, W.~M., {Tacchella}, S., {Johnson}, B.~D., {et~al.} 2025{\natexlab{c}}, Nature Astronomy, 9, 141

\bibitem[{{Baker} {et~al.}(2025{\natexlab{d}}){Baker}, {Valentino}, {Lagos}, {Ito}, {Jespersen}, {Gottumukkala}, {Hjorth}, {Langeroodi}, \& {Sedgewick}}]{Baker2025d}
{Baker}, W.~M., {Valentino}, F., {Lagos}, C. d.~P., {et~al.} 2025{\natexlab{d}}, \aap, 702, A270

\bibitem[{{Barrufet} {et~al.}(2025){Barrufet}, {Oesch}, {Marques-Chaves}, {Arellano-Cordova}, {Baggen}, {Carnall}, {Cullen}, {Dunlop}, {Gottumukkala}, {Fudamoto}, {Illingworth}, {Magee}, {McLure}, {McLeod}, {Micha{\l}owski}, {Stefanon}, {van Dokkum}, \& {Weibel}}]{Barrufet2025}
{Barrufet}, L., {Oesch}, P.~A., {Marques-Chaves}, R., {et~al.} 2025, \mnras, 537, 3453

\bibitem[{{Belli} {et~al.}(2019){Belli}, {Newman}, \& {Ellis}}]{Belli2019}
{Belli}, S., {Newman}, A.~B., \& {Ellis}, R.~S. 2019, \apj, 874, 17

\bibitem[{{Brammer}(2023)}]{Brammer2023}
{Brammer}, G. 2023, {grizli}

\bibitem[{{Brammer} {et~al.}(2008){Brammer}, {van Dokkum}, \& {Coppi}}]{Brammer2008}
{Brammer}, G.~B., {van Dokkum}, P.~G., \& {Coppi}, P. 2008, \apj, 686, 1503

\bibitem[{{Bruzual} \& {Charlot}(2003)}]{Bruzual2003}
{Bruzual}, G. \& {Charlot}, S. 2003, \mnras, 344, 1000

\bibitem[{{Calzetti} {et~al.}(2000){Calzetti}, {Armus}, {Bohlin}, {Kinney}, {Koornneef}, \& {Storchi-Bergmann}}]{Calzetti2000}
{Calzetti}, D., {Armus}, L., {Bohlin}, R.~C., {et~al.} 2000, \apj, 533, 682

\bibitem[{{Carnall} {et~al.}(2024){Carnall}, {Cullen}, {McLure}, {McLeod}, {Begley}, {Donnan}, {Dunlop}, {Shapley}, {Rowlands}, {Almaini}, {Arellano-C{\'o}rdova}, {Barrufet}, {Cimatti}, {Ellis}, {Grogin}, {Hamadouche}, {Illingworth}, {Koekemoer}, {Leung}, {Lovell}, {P{\'e}rez-Gonz{\'a}lez}, {Santini}, {Stanton}, \& {Wild}}]{Carnall2024}
{Carnall}, A.~C., {Cullen}, F., {McLure}, R.~J., {et~al.} 2024, \mnras, 534, 325

\bibitem[{{Carnall} {et~al.}(2018){Carnall}, {McLure}, {Dunlop}, \& {Dav{\'e}}}]{Carnall2018}
{Carnall}, A.~C., {McLure}, R.~J., {Dunlop}, J.~S., \& {Dav{\'e}}, R. 2018, \mnras, 480, 4379

\bibitem[{{Carnall} {et~al.}(2023){Carnall}, {McLure}, {Dunlop}, {McLeod}, {Wild}, {Cullen}, {Magee}, {Begley}, {Cimatti}, {Donnan}, {Hamadouche}, {Jewell}, \& {Walker}}]{Carnall2023Nature}
{Carnall}, A.~C., {McLure}, R.~J., {Dunlop}, J.~S., {et~al.} 2023, \nat, 619, 716

\bibitem[{{Carniani} {et~al.}(2025){Carniani}, {D'Eugenio}, {Ji}, {Parlanti}, {Scholtz}, {Sun}, {Venturi}, {Bakx}, {Curti}, {Maiolino}, {Tacchella}, {Zavala}, {Hainline}, {Witstok}, {Johnson}, {Alberts}, {Bunker}, {Charlot}, {Eisenstein}, {Helton}, {Jakobsen}, {Kumari}, {Robertson}, {Saxena}, {{\"U}bler}, {Williams}, {Willmer}, \& {Willott}}]{Carniani2025}
{Carniani}, S., {D'Eugenio}, F., {Ji}, X., {et~al.} 2025, \aap, 696, A87

\bibitem[{{Casey} {et~al.}(2014){Casey}, {Narayanan}, \& {Cooray}}]{Casey2014}
{Casey}, C.~M., {Narayanan}, D., \& {Cooray}, A. 2014, \physrep, 541, 45

\bibitem[{{Castellano} {et~al.}(2024){Castellano}, {Napolitano}, {Fontana}, {Roberts-Borsani}, {Treu}, {Vanzella}, {Zavala}, {Arrabal Haro}, {Calabr{\`o}}, {Llerena}, {Mascia}, {Merlin}, {Paris}, {Pentericci}, {Santini}, {Bakx}, {Bergamini}, {Cupani}, {Dickinson}, {Filippenko}, {Glazebrook}, {Grillo}, {Kelly}, {Malkan}, {Mason}, {Morishita}, {Nanayakkara}, {Rosati}, {Sani}, {Wang}, \& {Yoon}}]{Castellano2024}
{Castellano}, M., {Napolitano}, L., {Fontana}, A., {et~al.} 2024, \apj, 972, 143

\bibitem[{{Chabrier}(2003)}]{Chabrier2003}
{Chabrier}, G. 2003, \pasp, 115, 763

\bibitem[{{Chittenden} {et~al.}(2025){Chittenden}, {Glazebrook}, {Nanayakkara}, {Kawinwanichakij}, {Lagos}, {Kimmig}, \& {Remus}}]{Chittenden2025}
{Chittenden}, H.~G., {Glazebrook}, K., {Nanayakkara}, T., {et~al.} 2025, arXiv e-prints, arXiv:2504.19696

\bibitem[{{Cochrane}(2025)}]{Cochrane2025}
{Cochrane}, R.~K. 2025, \mnras, 544, 1530

\bibitem[{{Conroy} \& {Gunn}(2010)}]{Conroy2010}
{Conroy}, C. \& {Gunn}, J.~E. 2010, \apj, 712, 833

\bibitem[{{Conroy} {et~al.}(2009){Conroy}, {Gunn}, \& {White}}]{Conroy2009}
{Conroy}, C., {Gunn}, J.~E., \& {White}, M. 2009, \apj, 699, 486

\bibitem[{{Covelo-Paz} {et~al.}(2025){Covelo-Paz}, {Meuwly}, {Oesch}, {Witten}, {Weibel}, {Carvajal-Bohorquez}, {Ciesla}, {Giovinazzo}, \& {Brammer}}]{Covelo-Paz2025}
{Covelo-Paz}, A., {Meuwly}, C., {Oesch}, P.~A., {et~al.} 2025, arXiv e-prints, arXiv:2506.22540

\bibitem[{{Croton} {et~al.}(2006){Croton}, {Springel}, {White}, {De Lucia}, {Frenk}, {Gao}, {Jenkins}, {Kauffmann}, {Navarro}, \& {Yoshida}}]{Croton2006}
{Croton}, D.~J., {Springel}, V., {White}, S. D.~M., {et~al.} 2006, \mnras, 365, 11

\bibitem[{{Curtis-Lake} {et~al.}(2023){Curtis-Lake}, {Bluck}, {d'Eugenio}, {Maiolino}, \& {Sijacki}}]{Curtis-Lake2023}
{Curtis-Lake}, E., {Bluck}, A., {d'Eugenio}, F., {Maiolino}, R., \& {Sijacki}, D. 2023, Nature Astronomy, 7, 247

\bibitem[{{Cutler} {et~al.}(2024){Cutler}, {Whitaker}, {Weaver}, {Wang}, {Pan}, {Bezanson}, {Furtak}, {Labbe}, {Leja}, {Price}, {Cheng}, {Clausen}, {Cullen}, {Dayal}, {de Graaff}, {Dickinson}, {Dunlop}, {Feldmann}, {Franx}, {Giavalisco}, {Glazebrook}, {Greene}, {Grogin}, {Illingworth}, {Koekemoer}, {Kokorev}, {Marchesini}, {Maseda}, {Miller}, {Nanayakkara}, {Nelson}, {Setton}, {Shipley}, \& {Suess}}]{Cutler2024}
{Cutler}, S.~E., {Whitaker}, K.~E., {Weaver}, J.~R., {et~al.} 2024, \apjl, 967, L23

\bibitem[{{de Graaff} {et~al.}(2025{\natexlab{a}}){de Graaff}, {Brammer}, {Weibel}, {Lewis}, {Maseda}, {Oesch}, {Bezanson}, {Boogaard}, {Cleri}, {Cooper}, {Gottumukkala}, {Greene}, {Hirschmann}, {Hviding}, {Katz}, {Labb{\'e}}, {Leja}, {Matthee}, {McConachie}, {Miller}, {Naidu}, {Price}, {Rix}, {Setton}, {Suess}, {Wang}, {Whitaker}, \& {Williams}}]{deGraaff2024}
{de Graaff}, A., {Brammer}, G., {Weibel}, A., {et~al.} 2025{\natexlab{a}}, \aap, 697, A189

\bibitem[{{de Graaff} {et~al.}(2025{\natexlab{b}}){de Graaff}, {Setton}, {Brammer}, {Cutler}, {Suess}, {Labb{\'e}}, {Leja}, {Weibel}, {Maseda}, {Whitaker}, {Bezanson}, {Boogaard}, {Cleri}, {De Lucia}, {Franx}, {Greene}, {Hirschmann}, {Matthee}, {McConachie}, {Naidu}, {Oesch}, {Price}, {Rix}, {Valentino}, {Wang}, \& {Williams}}]{deGraaff2025}
{de Graaff}, A., {Setton}, D.~J., {Brammer}, G., {et~al.} 2025{\natexlab{b}}, Nature Astronomy, 9, 280

\bibitem[{{Dome} {et~al.}(2024){Dome}, {Tacchella}, {Fialkov}, {Ceverino}, {Dekel}, {Ginzburg}, {Lapiner}, \& {Looser}}]{Dome2024}
{Dome}, T., {Tacchella}, S., {Fialkov}, A., {et~al.} 2024, \mnras, 527, 2139

\bibitem[{{Donnan} {et~al.}(2024){Donnan}, {McLure}, {Dunlop}, {McLeod}, {Magee}, {Arellano-C{\'o}rdova}, {Barrufet}, {Begley}, {Bowler}, {Carnall}, {Cullen}, {Ellis}, {Fontana}, {Illingworth}, {Grogin}, {Hamadouche}, {Koekemoer}, {Liu}, {Mason}, {Santini}, \& {Stanton}}]{Donnan2024}
{Donnan}, C.~T., {McLure}, R.~J., {Dunlop}, J.~S., {et~al.} 2024, \mnras, 533, 3222

\bibitem[{{Dressler}(1980)}]{Dressler1980}
{Dressler}, A. 1980, \apj, 236, 351

\bibitem[{{Eisenstein} {et~al.}(2023){Eisenstein}, {Willott}, {Alberts}, {Arribas}, {Bonaventura}, {Bunker}, {Cameron}, {Carniani}, {Charlot}, {Curtis-Lake}, {D'Eugenio}, {Endsley}, {Ferruit}, {Giardino}, {Hainline}, {Hausen}, {Jakobsen}, {Johnson}, {Maiolino}, {Rieke}, {Rieke}, {Rix}, {Robertson}, {Stark}, {Tacchella}, {Williams}, {Willmer}, {Baker}, {Baum}, {Bhatawdekar}, {Boyett}, {Chen}, {Chevallard}, {Circosta}, {Curti}, {Danhaive}, {DeCoursey}, {de Graaff}, {Dressler}, {Egami}, {Helton}, {Hviding}, {Ji}, {Jones}, {Kumari}, {L{\"u}tzgendorf}, {Laseter}, {Looser}, {Lyu}, {Maseda}, {Nelson}, {Parlanti}, {Perna}, {Pusk{\'a}s}, {Rawle}, {Rodr{\'\i}guez Del Pino}, {Sandles}, {Saxena}, {Scholtz}, {Sharpe}, {Shivaei}, {Silcock}, {Simmonds}, {Skarbinski}, {Smit}, {Stone}, {Suess}, {Sun}, {Tang}, {Topping}, {{\"U}bler}, {Villanueva}, {Wallace}, {Whitler}, {Witstok}, \& {Woodrum}}]{Eisenstein2023}
{Eisenstein}, D.~J., {Willott}, C., {Alberts}, S., {et~al.} 2023, arXiv e-prints, arXiv:2306.02465

\bibitem[{{Endsley} {et~al.}(2025){Endsley}, {Chisholm}, {Stark}, {Topping}, \& {Whitler}}]{Endsley2025}
{Endsley}, R., {Chisholm}, J., {Stark}, D.~P., {Topping}, M.~W., \& {Whitler}, L. 2025, \apj, 987, 189

\bibitem[{{Endsley} {et~al.}(2024){Endsley}, {Stark}, {Whitler}, {Topping}, {Johnson}, {Robertson}, {Tacchella}, {Alberts}, {Baker}, {Bhatawdekar}, {Boyett}, {Bunker}, {Cameron}, {Carniani}, {Charlot}, {Chen}, {Chevallard}, {Curtis-Lake}, {Danhaive}, {Egami}, {Eisenstein}, {Hainline}, {Helton}, {Ji}, {Looser}, {Maiolino}, {Nelson}, {Pusk{\'a}s}, {Rieke}, {Rieke}, {Rix}, {Sandles}, {Saxena}, {Simmonds}, {Smit}, {Sun}, {Williams}, {Willmer}, {Willott}, \& {Witstok}}]{Endsley2024}
{Endsley}, R., {Stark}, D.~P., {Whitler}, L., {et~al.} 2024, \mnras, 533, 1111

\bibitem[{{Fabian}(2012)}]{Fabian2012}
{Fabian}, A.~C. 2012, \araa, 50, 455

\bibitem[{{Franx} {et~al.}(2008){Franx}, {van Dokkum}, {F{\"o}rster Schreiber}, {Wuyts}, {Labb{\'e}}, \& {Toft}}]{Franx2008}
{Franx}, M., {van Dokkum}, P.~G., {F{\"o}rster Schreiber}, N.~M., {et~al.} 2008, \apj, 688, 770

\bibitem[{{Fudamoto} {et~al.}(2025){Fudamoto}, {Helton}, {Lin}, {Sun}, {Behroozi}, {Hsiao}, {Egami}, {Bunker}, {Harikane}, {Ouchi}, {Liu}, {Liu}, {Maiolino}, {Ji}, {Jin}, {Tee}, {Wang}, {Willmer}, {Xu}, \& {Zhu}}]{Fudamoto2025}
{Fudamoto}, Y., {Helton}, J.~M., {Lin}, X., {et~al.} 2025, arXiv e-prints, arXiv:2503.15597

\bibitem[{{Gallazzi} {et~al.}(2014){Gallazzi}, {Bell}, {Zibetti}, {Brinchmann}, \& {Kelson}}]{Gallazzi2014}
{Gallazzi}, A., {Bell}, E.~F., {Zibetti}, S., {Brinchmann}, J., \& {Kelson}, D.~D. 2014, \apj, 788, 72

\bibitem[{{Gelli} {et~al.}(2025){Gelli}, {Pallottini}, {Salvadori}, {Ferrara}, {Mason}, {Carniani}, \& {Ginolfi}}]{Gelli2025}
{Gelli}, V., {Pallottini}, A., {Salvadori}, S., {et~al.} 2025, \apj, 985, 126

\bibitem[{{Genin} {et~al.}(2025){Genin}, {Shuntov}, {Brammer}, {Allen}, {Ito}, {Magdis}, {Matharu}, {Oesch}, {Toft}, \& {Valentino}}]{Genin2025}
{Genin}, A., {Shuntov}, M., {Brammer}, G., {et~al.} 2025, \aap, 699, A343

\bibitem[{{Glazebrook} {et~al.}(2024){Glazebrook}, {Nanayakkara}, {Schreiber}, {Lagos}, {Kawinwanichakij}, {Jacobs}, {Chittenden}, {Brammer}, {Kacprzak}, {Labbe}, {Marchesini}, {Marsan}, {Oesch}, {Papovich}, {Remus}, {Tran}, {Esdaile}, \& {Chandro-Gomez}}]{Glazebrook2024}
{Glazebrook}, K., {Nanayakkara}, T., {Schreiber}, C., {et~al.} 2024, \nat, 628, 277

\bibitem[{{Glazebrook} {et~al.}(2017){Glazebrook}, {Schreiber}, {Labb{\'e}}, {Nanayakkara}, {Kacprzak}, {Oesch}, {Papovich}, {Spitler}, {Straatman}, {Tran}, \& {Yuan}}]{Glazebrook2017}
{Glazebrook}, K., {Schreiber}, C., {Labb{\'e}}, I., {et~al.} 2017, \nat, 544, 71

\bibitem[{{Gottumukkala} {et~al.}(2024){Gottumukkala}, {Barrufet}, {Oesch}, {Weibel}, {Allen}, {Alcalde Pampliega}, {Nelson}, {Williams}, {Brammer}, {Fudamoto}, {Gonz{\'a}lez}, {Heintz}, {Illingworth}, {Magee}, {Naidu}, {Shuntov}, {Stefanon}, {Toft}, {Valentino}, \& {Xiao}}]{Gottumukkala2024}
{Gottumukkala}, R., {Barrufet}, L., {Oesch}, P.~A., {et~al.} 2024, \mnras, 530, 966

\bibitem[{{Grogin} {et~al.}(2011){Grogin}, {Kocevski}, {Faber}, {Ferguson}, {Koekemoer}, {Riess}, {Acquaviva}, {Alexander}, {Almaini}, {Ashby}, {Barden}, {Bell}, {Bournaud}, {Brown}, {Caputi}, {Casertano}, {Cassata}, {Castellano}, {Challis}, {Chary}, {Cheung}, {Cirasuolo}, {Conselice}, {Roshan Cooray}, {Croton}, {Daddi}, {Dahlen}, {Dav{\'e}}, {de Mello}, {Dekel}, {Dickinson}, {Dolch}, {Donley}, {Dunlop}, {Dutton}, {Elbaz}, {Fazio}, {Filippenko}, {Finkelstein}, {Fontana}, {Gardner}, {Garnavich}, {Gawiser}, {Giavalisco}, {Grazian}, {Guo}, {Hathi}, {H{\"a}ussler}, {Hopkins}, {Huang}, {Huang}, {Jha}, {Kartaltepe}, {Kirshner}, {Koo}, {Lai}, {Lee}, {Li}, {Lotz}, {Lucas}, {Madau}, {McCarthy}, {McGrath}, {McIntosh}, {McLure}, {Mobasher}, {Moustakas}, {Mozena}, {Nandra}, {Newman}, {Niemi}, {Noeske}, {Papovich}, {Pentericci}, {Pope}, {Primack}, {Rajan}, {Ravindranath}, {Reddy}, {Renzini}, {Rix}, {Robaina}, {Rodney}, {Rosario}, {Rosati}, {Salimbeni}, {Scarlata}, {Siana}, {Simard}, {Smidt}, {Somerville}, {Spinrad},
  {Straughn}, {Strolger}, {Telford}, {Teplitz}, {Trump}, {van der Wel}, {Villforth}, {Wechsler}, {Weiner}, {Wiklind}, {Wild}, {Wilson}, {Wuyts}, {Yan}, \& {Yun}}]{Grogin2011}
{Grogin}, N.~A., {Kocevski}, D.~D., {Faber}, S.~M., {et~al.} 2011, \apjs, 197, 35

\bibitem[{{Hamadouche} {et~al.}(2025){Hamadouche}, {McLure}, {Carnall}, {McLeod}, {Dunlop}, {Whitaker}, {Donnan}, {Begley}, {Stanton}, {Almaini}, {Aird}, {Cullen}, {Cutler}, {Grogin}, \& {Koekemoer}}]{Hamadouche2024}
{Hamadouche}, M.~L., {McLure}, R.~J., {Carnall}, A.~C., {et~al.} 2025, \mnras, 541, 463

\bibitem[{Harris {et~al.}(2020)Harris, Millman, van~der Walt, Gommers, Virtanen, Cournapeau, Wieser, Taylor, Berg, Smith, Kern, Picus, Hoyer, van Kerkwijk, Brett, Haldane, del R{\'{i}}o, Wiebe, Peterson, G{\'{e}}rard-Marchant, Sheppard, Reddy, Weckesser, Abbasi, Gohlke, \& Oliphant}]{harris2020array}
Harris, C.~R., Millman, K.~J., van~der Walt, S.~J., {et~al.} 2020, Nature, 585, 357

\bibitem[{{Heintz} {et~al.}(2025){Heintz}, {Brammer}, {Watson}, {Oesch}, {Keating}, {Hayes}, {Abdurro'uf}, {Arellano-C{\'o}rdova}, {Carnall}, {Christiansen}, {Cullen}, {Dav{\'e}}, {Dayal}, {Ferrara}, {Finlator}, {Fynbo}, {Flury}, {Gelli}, {Gillman}, {Gottumukkala}, {Gould}, {Greve}, {Hardin}, {Hsiao}, {Hutter}, {Jakobsson}, {Killi}, {Khosravaninezhad}, {Laursen}, {Lee}, {Magdis}, {Matthee}, {Naidu}, {Narayanan}, {Pollock}, {Prescott}, {Rusakov}, {Shuntov}, {Sneppen}, {Smit}, {Tanvir}, {Terp}, {Toft}, {Valentino}, {Vijayan}, {Weaver}, {Wise}, \& {Witstok}}]{Heintz2025}
{Heintz}, K.~E., {Brammer}, G.~B., {Watson}, D., {et~al.} 2025, \aap, 693, A60

\bibitem[{{Helton} {et~al.}(2024{\natexlab{a}}){Helton}, {Sun}, {Woodrum}, {Hainline}, {Willmer}, {Rieke}, {Rieke}, {Tacchella}, {Robertson}, {Johnson}, {Alberts}, {Eisenstein}, {Hausen}, {Bonaventura}, {Bunker}, {Charlot}, {Curti}, {Curtis-Lake}, {Looser}, {Maiolino}, {Willott}, {Witstok}, {Boyett}, {Chen}, {Egami}, {Endsley}, {Hviding}, {Jaffe}, {Ji}, {Lyu}, \& {Sandles}}]{Helton2024}
{Helton}, J.~M., {Sun}, F., {Woodrum}, C., {et~al.} 2024{\natexlab{a}}, \apj, 962, 124

\bibitem[{{Helton} {et~al.}(2024{\natexlab{b}}){Helton}, {Sun}, {Woodrum}, {Hainline}, {Willmer}, {Rieke}, {Rieke}, {Alberts}, {Eisenstein}, {Tacchella}, {Robertson}, {Johnson}, {Baker}, {Bhatawdekar}, {Bunker}, {Chen}, {Egami}, {Ji}, {Maiolino}, {Willott}, \& {Witstok}}]{Helton2024b}
{Helton}, J.~M., {Sun}, F., {Woodrum}, C., {et~al.} 2024{\natexlab{b}}, \apj, 974, 41

\bibitem[{Hunter(2007)}]{Hunter:2007}
Hunter, J.~D. 2007, Computing in Science \& Engineering, 9, 90

\bibitem[{{Ito} {et~al.}(2024){Ito}, {Valentino}, {Brammer}, {Faisst}, {Gillman}, {G{\'o}mez-Guijarro}, {Gould}, {Heintz}, {Ilbert}, {Jespersen}, {Kokorev}, {Kubo}, {Magdis}, {McPartland}, {Onodera}, {Rizzo}, {Tanaka}, {Toft}, {Vijayan}, {Weaver}, {Whitaker}, \& {Wright}}]{Ito2024}
{Ito}, K., {Valentino}, F., {Brammer}, G., {et~al.} 2024, \apj, 964, 192

\bibitem[{{Ito} {et~al.}(2025{\natexlab{a}}){Ito}, {Valentino}, {Brammer}, {Hamadouche}, {Whitaker}, {Kokorev}, {Zhu}, {Kakimoto}, {Wu}, {Antwi-Danso}, {Baker}, {Ceverino}, {Faisst}, {Farcy}, {Fujimoto}, {Gallazzi}, {Gillman}, {Gottumukkala}, {Heintz}, {Hirschmann}, {Jespersen}, {Kubo}, {Lee}, {Magdis}, {Onodera}, {Shimakawa}, {Tanaka}, {Toft}, \& {Weaver}}]{Ito2025b}
{Ito}, K., {Valentino}, F., {Brammer}, G., {et~al.} 2025{\natexlab{a}}, arXiv e-prints, arXiv:2506.22642

\bibitem[{{Ito} {et~al.}(2025{\natexlab{b}}){Ito}, {Valentino}, {Farcy}, {De Lucia}, {Lagos}, {Hirschmann}, {Brammer}, {de Graaff}, {Bl{\'a}nquez-Ses{\'e}}, {Ceverino}, {Faisst}, {Fontanot}, {Gillman}, {Hamadouche}, {Heintz}, {Jin}, {Jespersen}, {Kubo}, {Lee}, {Magdis}, {Man}, {Onodera}, {Rizzo}, {Shimakawa}, {Tanaka}, {Toft}, {Whitaker}, {Xie}, \& {Zhu}}]{Ito2025}
{Ito}, K., {Valentino}, F., {Farcy}, M., {et~al.} 2025{\natexlab{b}}, \aap, 697, A111

\bibitem[{{Jespersen} {et~al.}(2025){Jespersen}, {Carnall}, \& {Lovell}}]{Jespersen2025b}
{Jespersen}, C.~K., {Carnall}, A.~C., \& {Lovell}, C.~C. 2025, \apjl, 988, L19

\bibitem[{{Johnson} {et~al.}(2021){Johnson}, {Leja}, {Conroy}, \& {Speagle}}]{Johnson2021}
{Johnson}, B.~D., {Leja}, J., {Conroy}, C., \& {Speagle}, J.~S. 2021, \apjs, 254, 22

\bibitem[{{Kakimoto} {et~al.}(2024){Kakimoto}, {Tanaka}, {Onodera}, {Shimakawa}, {Wu}, {Gould}, {Ito}, {Jin}, {Kubo}, {Suzuki}, {Toft}, {Valentino}, \& {Yabe}}]{Kakimoto2024}
{Kakimoto}, T., {Tanaka}, M., {Onodera}, M., {et~al.} 2024, \apj, 963, 49

\bibitem[{{Kawinwanichakij} {et~al.}(2025){Kawinwanichakij}, {Glazebrook}, {Nanayakkara}, {Kacprzak}, {Chittenden}, {Jacobs}, {Chandro-G{\'o}mez}, {Lagos}, {Marchesini}, {Mart{\'\i}nez-Mar{\'\i}n}, {Oesch}, \& {Remus}}]{Kawinwanichakij2025}
{Kawinwanichakij}, L., {Glazebrook}, K., {Nanayakkara}, T., {et~al.} 2025, arXiv e-prints, arXiv:2505.03089

\bibitem[{{Kawinwanichakij} {et~al.}(2017){Kawinwanichakij}, {Papovich}, {Quadri}, {Glazebrook}, {Kacprzak}, {Allen}, {Bell}, {Croton}, {Dekel}, {Ferguson}, {Forrest}, {Grogin}, {Guo}, {Kocevski}, {Koekemoer}, {Labb{\'e}}, {Lucas}, {Nanayakkara}, {Spitler}, {Straatman}, {Tran}, {Tomczak}, \& {van Dokkum}}]{Kawinwanichakij2017}
{Kawinwanichakij}, L., {Papovich}, C., {Quadri}, R.~F., {et~al.} 2017, \apj, 847, 134

\bibitem[{{Kennicutt} \& {Evans}(2012)}]{Kennicutt2012}
{Kennicutt}, R.~C. \& {Evans}, N.~J. 2012, \araa, 50, 531

\bibitem[{{Kennicutt}(1998)}]{Kennicutt1998}
{Kennicutt}, Jr., R.~C. 1998, \araa, 36, 189

\bibitem[{{Koekemoer} {et~al.}(2011){Koekemoer}, {Faber}, {Ferguson}, {Grogin}, {Kocevski}, {Koo}, {Lai}, {Lotz}, {Lucas}, {McGrath}, {Ogaz}, {Rajan}, {Riess}, {Rodney}, {Strolger}, {Casertano}, {Castellano}, {Dahlen}, {Dickinson}, {Dolch}, {Fontana}, {Giavalisco}, {Grazian}, {Guo}, {Hathi}, {Huang}, {van der Wel}, {Yan}, {Acquaviva}, {Alexander}, {Almaini}, {Ashby}, {Barden}, {Bell}, {Bournaud}, {Brown}, {Caputi}, {Cassata}, {Challis}, {Chary}, {Cheung}, {Cirasuolo}, {Conselice}, {Roshan Cooray}, {Croton}, {Daddi}, {Dav{\'e}}, {de Mello}, {de Ravel}, {Dekel}, {Donley}, {Dunlop}, {Dutton}, {Elbaz}, {Fazio}, {Filippenko}, {Finkelstein}, {Frazer}, {Gardner}, {Garnavich}, {Gawiser}, {Gruetzbauch}, {Hartley}, {H{\"a}ussler}, {Herrington}, {Hopkins}, {Huang}, {Jha}, {Johnson}, {Kartaltepe}, {Khostovan}, {Kirshner}, {Lani}, {Lee}, {Li}, {Madau}, {McCarthy}, {McIntosh}, {McLure}, {McPartland}, {Mobasher}, {Moreira}, {Mortlock}, {Moustakas}, {Mozena}, {Nandra}, {Newman}, {Nielsen}, {Niemi}, {Noeske}, {Papovich},
  {Pentericci}, {Pope}, {Primack}, {Ravindranath}, {Reddy}, {Renzini}, {Rix}, {Robaina}, {Rosario}, {Rosati}, {Salimbeni}, {Scarlata}, {Siana}, {Simard}, {Smidt}, {Snyder}, {Somerville}, {Spinrad}, {Straughn}, {Telford}, {Teplitz}, {Trump}, {Vargas}, {Villforth}, {Wagner}, {Wandro}, {Wechsler}, {Weiner}, {Wiklind}, {Wild}, {Wilson}, {Wuyts}, \& {Yun}}]{Koekemoer2011}
{Koekemoer}, A.~M., {Faber}, S.~M., {Ferguson}, H.~C., {et~al.} 2011, \apjs, 197, 36

\bibitem[{{Kokorev} {et~al.}(2025){Kokorev}, {Ch{\'a}vez Ortiz}, {Taylor}, {Finkelstein}, {Arrabal Haro}, {Dickinson}, {Chisholm}, {Fujimoto}, {noz}, {Endsley}, {Hu}, {Napolitano}, {Wilkins}, {Akins}, {Amori{\'\i}n}, {Casey}, {Cheng}, {Cleri}, {Cole}, {Cullen}, {Daddi}, {Davis}, {Donnan}, {Dunlop}, {Fern{\'a}ndez}, {Giavalisco}, {Grogin}, {Hathi}, {Hirschmann}, {Kartaltepe}, {Koekemoer}, {Leung}, {Lucas}, {McLeod}, {Papovich}, {Pentericci}, {P{\'e}rez-Gonz{\'a}lez}, {Somerville}, {Wang}, {Yung}, \& {Zavala}}]{Kokorev2025_capers}
{Kokorev}, V., {Ch{\'a}vez Ortiz}, {\'O}.~A., {Taylor}, A.~J., {et~al.} 2025, \apjl, 988, L10

\bibitem[{{Kron}(1980)}]{Kron1980}
{Kron}, R.~G. 1980, \apjs, 43, 305

\bibitem[{{Langeroodi} \& {Hjorth}(2024)}]{Langeroodi2024}
{Langeroodi}, D. \& {Hjorth}, J. 2024, arXiv e-prints, arXiv:2404.13045

\bibitem[{{Laporte} {et~al.}(2022){Laporte}, {Zitrin}, {Dole}, {Roberts-Borsani}, {Furtak}, \& {Witten}}]{Laporte2022}
{Laporte}, N., {Zitrin}, A., {Dole}, H., {et~al.} 2022, \aap, 667, L3

\bibitem[{{Leja} {et~al.}(2019{\natexlab{a}}){Leja}, {Carnall}, {Johnson}, {Conroy}, \& {Speagle}}]{Leja2019}
{Leja}, J., {Carnall}, A.~C., {Johnson}, B.~D., {Conroy}, C., \& {Speagle}, J.~S. 2019{\natexlab{a}}, \apj, 876, 3

\bibitem[{{Leja} {et~al.}(2019{\natexlab{b}}){Leja}, {Johnson}, {Conroy}, {van Dokkum}, {Speagle}, {Brammer}, {Momcheva}, {Skelton}, {Whitaker}, {Franx}, \& {Nelson}}]{Leja2019stellarmasses}
{Leja}, J., {Johnson}, B.~D., {Conroy}, C., {et~al.} 2019{\natexlab{b}}, \apj, 877, 140

\bibitem[{{Le{\'s}niewska} {et~al.}(2025){Le{\'s}niewska}, {Hjorth}, \& {Gall}}]{Lesniewska2025}
{Le{\'s}niewska}, A., {Hjorth}, J., \& {Gall}, C. 2025, \aap, 699, A352

\bibitem[{{Looser} {et~al.}(2025){Looser}, {D'Eugenio}, {Maiolino}, {Tacchella}, {Curti}, {Arribas}, {Baker}, {Baum}, {Bonaventura}, {Boyett}, {Bunker}, {Carniani}, {Charlot}, {Chevallard}, {Curtis-Lake}, {Lola Danhaive}, {Eisenstein}, {de Graaff}, {Hainline}, {Ji}, {Johnson}, {Kumari}, {Nelson}, {Parlanti}, {Rix}, {Robertson}, {Del Pino}, {Sandles}, {Scholtz}, {Smit}, {Stark}, {{\"U}bler}, {Williams}, {Willott}, \& {Witstok}}]{Looser2025}
{Looser}, T.~J., {D'Eugenio}, F., {Maiolino}, R., {et~al.} 2025, \aap, 697, A88

\bibitem[{{Looser} {et~al.}(2024){Looser}, {D'Eugenio}, {Maiolino}, {Witstok}, {Sandles}, {Curtis-Lake}, {Chevallard}, {Tacchella}, {Johnson}, {Baker}, {Suess}, {Carniani}, {Ferruit}, {Arribas}, {Bonaventura}, {Bunker}, {Cameron}, {Charlot}, {Curti}, {de Graaff}, {Maseda}, {Rawle}, {Rix}, {Del Pino}, {Smit}, {{\"U}bler}, {Willott}, {Alberts}, {Egami}, {Eisenstein}, {Endsley}, {Hausen}, {Rieke}, {Robertson}, {Shivaei}, {Williams}, {Boyett}, {Chen}, {Ji}, {Jones}, {Kumari}, {Nelson}, {Perna}, {Saxena}, \& {Scholtz}}]{Looser2024}
{Looser}, T.~J., {D'Eugenio}, F., {Maiolino}, R., {et~al.} 2024, \nat, 629, 53

\bibitem[{{Man} \& {Belli}(2018)}]{Man2018}
{Man}, A. \& {Belli}, S. 2018, Nature Astronomy, 2, 695

\bibitem[{{McConachie} {et~al.}(2025){McConachie}, {de Graaff}, {Maseda}, {Leja}, {Zhang}, {Setton}, {Bezanson}, {Boogaard}, {Brammer}, {Cleri}, {Cooper}, {Glazebrook}, {Gottumukkala}, {Greene}, {Goulding}, {Hirschmann}, {Labbe}, {Lewis}, {Matthee}, {Miller}, {Naidu}, {Oesch}, {Price}, {Nanayakkara}, {Suess}, {Wang}, {Whitaker}, \& {Williams}}]{McConachie2025}
{McConachie}, I., {de Graaff}, A., {Maseda}, M.~V., {et~al.} 2025, arXiv e-prints, arXiv:2510.25024

\bibitem[{{McConachie} {et~al.}(2022){McConachie}, {Wilson}, {Forrest}, {Marsan}, {Muzzin}, {Cooper}, {Annunziatella}, {Marchesini}, {Chan}, {Gomez}, {Abdullah}, {Saracco}, \& {Nantais}}]{McConachie2022}
{McConachie}, I., {Wilson}, G., {Forrest}, B., {et~al.} 2022, \apj, 926, 37

\bibitem[{{Moore} {et~al.}(1996){Moore}, {Katz}, {Lake}, {Dressler}, \& {Oemler}}]{Moore1996}
{Moore}, B., {Katz}, N., {Lake}, G., {Dressler}, A., \& {Oemler}, A. 1996, \nat, 379, 613

\bibitem[{{Morishita} {et~al.}(2025){Morishita}, {Liu}, {Stiavelli}, {Treu}, {Trenti}, {Chartab}, {Roberts-Borsani}, {Vulcani}, {Bergamini}, {Castellano}, \& {Grillo}}]{Morishita2025}
{Morishita}, T., {Liu}, Z., {Stiavelli}, M., {et~al.} 2025, \apj, 982, 153

\bibitem[{{Naidu} {et~al.}(2025){Naidu}, {Oesch}, {Brammer}, {Weibel}, {Li}, {Matthee}, {Chisholm}, {Pollock}, {Heintz}, {Johnson}, {Shen}, {Hviding}, {Leja}, {Tacchella}, {Ganguly}, {Witten}, {Atek}, {Belli}, {Bose}, {Bouwens}, {Dayal}, {Decarli}, {de Graaff}, {Fudamoto}, {Giovinazzo}, {Greene}, {Illingworth}, {Inoue}, {Kane}, {Labbe}, {Leonova}, {Marques-Chaves}, {Meyer}, {Nelson}, {Roberts-Borsani}, {Schaerer}, {Simcoe}, {Stefanon}, {Sugahara}, {Toft}, {van der Wel}, {van Dokkum}, {Walter}, {Watson}, {Weaver}, \& {Whitaker}}]{Naidu2025}
{Naidu}, R.~P., {Oesch}, P.~A., {Brammer}, G., {et~al.} 2025, arXiv e-prints, arXiv:2505.11263

\bibitem[{{Nanayakkara} {et~al.}(2025){Nanayakkara}, {Glazebrook}, {Schreiber}, {Chittenden}, {Brammer}, {Esdaile}, {Jacobs}, {Kacprzak}, {Kawinwanichakij}, {Kimmig}, {Labbe}, {Lagos}, {Marchesini}, {Mart{\`\i}nez-Mar{\`\i}n}, {Marsan}, {Oesch}, {Papovich}, {Remus}, \& {Tran}}]{Nanayakkara2025}
{Nanayakkara}, T., {Glazebrook}, K., {Schreiber}, C., {et~al.} 2025, \apj, 981, 78

\bibitem[{{Overzier}(2016)}]{Overzier2016}
{Overzier}, R.~A. 2016, \aapr, 24, 14

\bibitem[{{Pasha} \& {Miller}(2023)}]{Pasha2023}
{Pasha}, I. \& {Miller}, T.~B. 2023, The Journal of Open Source Software, 8, 5703

\bibitem[{{Peng} {et~al.}(2010){Peng}, {Lilly}, {Kova{\v{c}}}, {Bolzonella}, {Pozzetti}, {Renzini}, {Zamorani}, {Ilbert}, {Knobel}, {Iovino}, {Maier}, {Cucciati}, {Tasca}, {Carollo}, {Silverman}, {Kampczyk}, {de Ravel}, {Sanders}, {Scoville}, {Contini}, {Mainieri}, {Scodeggio}, {Kneib}, {Le F{\`e}vre}, {Bardelli}, {Bongiorno}, {Caputi}, {Coppa}, {de la Torre}, {Franzetti}, {Garilli}, {Lamareille}, {Le Borgne}, {Le Brun}, {Mignoli}, {Perez Montero}, {Pello}, {Ricciardelli}, {Tanaka}, {Tresse}, {Vergani}, {Welikala}, {Zucca}, {Oesch}, {Abbas}, {Barnes}, {Bordoloi}, {Bottini}, {Cappi}, {Cassata}, {Cimatti}, {Fumana}, {Hasinger}, {Koekemoer}, {Leauthaud}, {Maccagni}, {Marinoni}, {McCracken}, {Memeo}, {Meneux}, {Nair}, {Porciani}, {Presotto}, \& {Scaramella}}]{Peng2010}
{Peng}, Y.-j., {Lilly}, S.~J., {Kova{\v{c}}}, K., {et~al.} 2010, \apj, 721, 193

\bibitem[{{Planck Collaboration} {et~al.}(2020){Planck Collaboration}, {Aghanim}, {Akrami}, {Ashdown}, {Aumont}, {Baccigalupi}, {Ballardini}, {Banday}, {Barreiro}, {Bartolo}, {Basak}, {Battye}, {Benabed}, {Bernard}, {Bersanelli}, {Bielewicz}, {Bock}, {Bond}, {Borrill}, {Bouchet}, {Boulanger}, {Bucher}, {Burigana}, {Butler}, {Calabrese}, {Cardoso}, {Carron}, {Challinor}, {Chiang}, {Chluba}, {Colombo}, {Combet}, {Contreras}, {Crill}, {Cuttaia}, {de Bernardis}, {de Zotti}, {Delabrouille}, {Delouis}, {Di Valentino}, {Diego}, {Dor{\'e}}, {Douspis}, {Ducout}, {Dupac}, {Dusini}, {Efstathiou}, {Elsner}, {En{\ss}lin}, {Eriksen}, {Fantaye}, {Farhang}, {Fergusson}, {Fernandez-Cobos}, {Finelli}, {Forastieri}, {Frailis}, {Fraisse}, {Franceschi}, {Frolov}, {Galeotta}, {Galli}, {Ganga}, {G{\'e}nova-Santos}, {Gerbino}, {Ghosh}, {Gonz{\'a}lez-Nuevo}, {G{\'o}rski}, {Gratton}, {Gruppuso}, {Gudmundsson}, {Hamann}, {Handley}, {Hansen}, {Herranz}, {Hildebrandt}, {Hivon}, {Huang}, {Jaffe}, {Jones}, {Karakci}, {Keih{\"a}nen},
  {Keskitalo}, {Kiiveri}, {Kim}, {Kisner}, {Knox}, {Krachmalnicoff}, {Kunz}, {Kurki-Suonio}, {Lagache}, {Lamarre}, {Lasenby}, {Lattanzi}, {Lawrence}, {Le Jeune}, {Lemos}, {Lesgourgues}, {Levrier}, {Lewis}, {Liguori}, {Lilje}, {Lilley}, {Lindholm}, {L{\'o}pez-Caniego}, {Lubin}, {Ma}, {Mac{\'\i}as-P{\'e}rez}, {Maggio}, {Maino}, {Mandolesi}, {Mangilli}, {Marcos-Caballero}, {Maris}, {Martin}, {Martinelli}, {Mart{\'\i}nez-Gonz{\'a}lez}, {Matarrese}, {Mauri}, {McEwen}, {Meinhold}, {Melchiorri}, {Mennella}, {Migliaccio}, {Millea}, {Mitra}, {Miville-Desch{\^e}nes}, {Molinari}, {Montier}, {Morgante}, {Moss}, {Natoli}, {N{\o}rgaard-Nielsen}, {Pagano}, {Paoletti}, {Partridge}, {Patanchon}, {Peiris}, {Perrotta}, {Pettorino}, {Piacentini}, {Polastri}, {Polenta}, {Puget}, {Rachen}, {Reinecke}, {Remazeilles}, {Renzi}, {Rocha}, {Rosset}, {Roudier}, {Rubi{\~n}o-Mart{\'\i}n}, {Ruiz-Granados}, {Salvati}, {Sandri}, {Savelainen}, {Scott}, {Shellard}, {Sirignano}, {Sirri}, {Spencer}, {Sunyaev}, {Suur-Uski}, {Tauber}, {Tavagnacco},
  {Tenti}, {Toffolatti}, {Tomasi}, {Trombetti}, {Valenziano}, {Valiviita}, {Van Tent}, {Vibert}, {Vielva}, {Villa}, {Vittorio}, {Wandelt}, {Wehus}, {White}, {White}, {Zacchei}, \& {Zonca}}]{PlanckCollaboration2020}
{Planck Collaboration}, {Aghanim}, N., {Akrami}, Y., {et~al.} 2020, \aap, 641, A6

\bibitem[{{Pollock} {et~al.}(2025){Pollock}, {Gottumukkala}, {Heintz}, {Brammer}, {Roberts-Borsani}, {Oesch}, {Witstok}, {Arellano-C{\'o}rdova}, {Cullen}, {Scholte}, {Terp}, {Rowland}, {Sneppen}, {Ito}, {Valentino}, {Matthee}, {Watson}, \& {Toft}}]{Pollock2025}
{Pollock}, C.~L., {Gottumukkala}, R., {Heintz}, K.~E., {et~al.} 2025, arXiv e-prints, arXiv:2506.15779

\bibitem[{{Pusk{\'a}s} {et~al.}(2025){Pusk{\'a}s}, {Tacchella}, {Simmonds}, {Hainline}, {D'Eugenio}, {Alberts}, {Arribas}, {Baker}, {Bunker}, {Carniani}, {Charlot}, {Duan}, {Eisenstein}, {Ji}, {Johnson}, {Jones}, {Maiolino}, {McClymont}, {Rieke}, {Rinaldi}, {Robertson}, {{\"U}bler}, {Williams}, {Willmer}, {Willott}, \& {Witstok}}]{Puskas2025}
{Pusk{\'a}s}, D., {Tacchella}, S., {Simmonds}, C., {et~al.} 2025, \mnras, 540, 2146

\bibitem[{{Rieke} {et~al.}(2023){Rieke}, {Robertson}, {Tacchella}, {Hainline}, {Johnson}, {Hausen}, {Ji}, {Willmer}, {Eisenstein}, {Pusk{\'a}s}, {Alberts}, {Arribas}, {Baker}, {Baum}, {Bhatawdekar}, {Bonaventura}, {Boyett}, {Bunker}, {Cameron}, {Carniani}, {Charlot}, {Chevallard}, {Chen}, {Curti}, {Curtis-Lake}, {Danhaive}, {DeCoursey}, {Dressler}, {Egami}, {Endsley}, {Helton}, {Hviding}, {Kumari}, {Looser}, {Lyu}, {Maiolino}, {Maseda}, {Nelson}, {Rieke}, {Rix}, {Sandles}, {Saxena}, {Sharpe}, {Shivaei}, {Skarbinski}, {Smit}, {Stark}, {Stone}, {Suess}, {Sun}, {Topping}, {{\"U}bler}, {Villanueva}, {Wallace}, {Williams}, {Willott}, {Whitler}, {Witstok}, \& {Woodrum}}]{Rieke2023}
{Rieke}, M.~J., {Robertson}, B., {Tacchella}, S., {et~al.} 2023, \apjs, 269, 16

\bibitem[{{Roberts-Borsani} {et~al.}(2024){Roberts-Borsani}, {Treu}, {Shapley}, {Fontana}, {Pentericci}, {Castellano}, {Morishita}, {Bergamini}, \& {Rosati}}]{Roberts-Borsani2024}
{Roberts-Borsani}, G., {Treu}, T., {Shapley}, A., {et~al.} 2024, \apj, 976, 193

\bibitem[{{Sandles} {et~al.}(2023){Sandles}, {D'Eugenio}, {Helton}, {Maiolino}, {Hainline}, {Baker}, {Williams}, {Alberts}, {Bunker}, {Carniani}, {Charlot}, {Chevallard}, {Curti}, {Curtis-Lake}, {Eisenstein}, {Ji}, {Johnson}, {Looser}, {Rawle}, {Robertson}, {Rodr{\'\i}guez Del Pino}, {Tacchella}, {{\"U}bler}, {Willmer}, \& {Willott}}]{Sandles2023}
{Sandles}, L., {D'Eugenio}, F., {Helton}, J.~M., {et~al.} 2023, arXiv e-prints, arXiv:2307.08633

\bibitem[{{Sato} {et~al.}(2024){Sato}, {Inoue}, {Harikane}, {Shimakawa}, {Sugahara}, {Tamura}, {Hashimoto}, {Ito}, {Yamanaka}, {Mawatari}, {Fudamoto}, \& {Ren}}]{Sato2024}
{Sato}, R.~A., {Inoue}, A.~K., {Harikane}, Y., {et~al.} 2024, \mnras, 534, 3552

\bibitem[{{Sawicki} \& {Yee}(1998)}]{Sawicki1998}
{Sawicki}, M. \& {Yee}, H.~K.~C. 1998, \aj, 115, 1329

\bibitem[{{Saxena} {et~al.}(2024){Saxena}, {Cameron}, {Katz}, {Bunker}, {Chevallard}, {D'Eugenio}, {Arribas}, {Bhatawdekar}, {Boyett}, {Cargile}, {Carniani}, {Charlot}, {Curti}, {Curtis-Lake}, {Hainline}, {Ji}, {Johnson}, {Jones}, {Kumari}, {Laseter}, {Maseda}, {Robertson}, {Simmonds}, {Tacchella}, {Ubler}, {Williams}, {Willott}, {Witstok}, \& {Zhu}}]{Saxena2024}
{Saxena}, A., {Cameron}, A.~J., {Katz}, H., {et~al.} 2024, arXiv e-prints, arXiv:2411.14532

\bibitem[{{Schreiber} {et~al.}(2015){Schreiber}, {Pannella}, {Elbaz}, {B{\'e}thermin}, {Inami}, {Dickinson}, {Magnelli}, {Wang}, {Aussel}, {Daddi}, {Juneau}, {Shu}, {Sargent}, {Buat}, {Faber}, {Ferguson}, {Giavalisco}, {Koekemoer}, {Magdis}, {Morrison}, {Papovich}, {Santini}, \& {Scott}}]{Schreiber2015}
{Schreiber}, C., {Pannella}, M., {Elbaz}, D., {et~al.} 2015, \aap, 575, A74

\bibitem[{{Sersic}(1968)}]{Sersic1968}
{Sersic}, J.~L. 1968, {Atlas de Galaxias Australes}

\bibitem[{{Simmonds} {et~al.}(2025){Simmonds}, {Tacchella}, {Curtis-Lake}, {D'Eugenio}, {Hainline}, {Johnson}, {Kravtsov}, {Pusk{\~A}{\textexclamdown}s}, {Robertson}, {Stoffers}, {Willott}, {Baker}, {Belokurov}, {Bhatawdekar}, {Bunker}, {Carniani}, {Chevallard}, {Curti}, {Duan}, {Helton}, {Ji}, {Looser}, {Maiolino}, {Maseda}, {Shivaei}, \& {Williams}}]{Simmonds2025}
{Simmonds}, C., {Tacchella}, S., {Curtis-Lake}, W. M.~E., {et~al.} 2025, \mnras [\eprint[arXiv]{2508.04410}]

\bibitem[{{Speagle}(2020)}]{Speagle2020}
{Speagle}, J.~S. 2020, \mnras, 493, 3132

\bibitem[{{Stevenson} {et~al.}(2025){Stevenson}, {Carnall}, {Leung}, {Taylor}, {Cullen}, {Dunlop}, {McLeod}, {McLure}, {Begley}, {Arellano-C{\'o}rdova}, {Barrufet}, {Bondestam}, {Donnan}, {Ellis}, {Grogin}, {Koekemoer}, {Liu}, {P{\'e}rez-Gonz{\'a}lez}, {Rowlands}, {Sanders}, {Scholte}, {Shapley}, {Skarbinski}, {Stanton}, \& {Wild}}]{Stevenson2025}
{Stevenson}, S.~D., {Carnall}, A.~C., {Leung}, H.-H., {et~al.} 2025, \mnras [\eprint[arXiv]{2509.06913}]

\bibitem[{{Strait} {et~al.}(2023){Strait}, {Brammer}, {Muzzin}, {Desprez}, {Asada}, {Abraham}, {Brada{\v{c}}}, {Iyer}, {Martis}, {Mowla}, {Noirot}, {Sarrouh}, {Sawicki}, {Willott}, {Gould}, {Grindlay}, {Matharu}, \& {Rihtar{\v{s}}i{\v{c}}}}]{Strait2023}
{Strait}, V., {Brammer}, G., {Muzzin}, A., {et~al.} 2023, \apjl, 949, L23

\bibitem[{{Tacchella} {et~al.}(2015){Tacchella}, {Carollo}, {Renzini}, {F{\"o}rster Schreiber}, {Lang}, {Wuyts}, {Cresci}, {Dekel}, {Genzel}, {Lilly}, {Mancini}, {Newman}, {Onodera}, {Shapley}, {Tacconi}, {Woo}, \& {Zamorani}}]{Tacchella2015}
{Tacchella}, S., {Carollo}, C.~M., {Renzini}, A., {et~al.} 2015, Science, 348, 314

\bibitem[{{Tacchella} {et~al.}(2022){Tacchella}, {Finkelstein}, {Bagley}, {Dickinson}, {Ferguson}, {Giavalisco}, {Graziani}, {Grogin}, {Hathi}, {Hutchison}, {Jung}, {Koekemoer}, {Larson}, {Papovich}, {Pirzkal}, {Rojas-Ruiz}, {Song}, {Schneider}, {Somerville}, {Wilkins}, \& {Yung}}]{Tacchella2022}
{Tacchella}, S., {Finkelstein}, S.~L., {Bagley}, M., {et~al.} 2022, \apj, 927, 170

\bibitem[{{Tanaka} {et~al.}(2024){Tanaka}, {Onodera}, {Shimakawa}, {Ito}, {Kakimoto}, {Kubo}, {Morishita}, {Toft}, {Valentino}, \& {Wu}}]{Tanaka2024}
{Tanaka}, M., {Onodera}, M., {Shimakawa}, R., {et~al.} 2024, \apj, 970, 59

\bibitem[{{Taylor}(2005)}]{Taylor2005}
{Taylor}, M.~B. 2005, in Astronomical Society of the Pacific Conference Series, Vol. 347, Astronomical Data Analysis Software and Systems XIV, ed. P.~{Shopbell}, M.~{Britton}, \& R.~{Ebert}, 29

\bibitem[{{Trussler} {et~al.}(2025){Trussler}, {Conselice}, {Adams}, {Austin}, {Caruana}, {Harvey}, {Li}, {Lovell}, {Seeyave}, {Vijayan}, \& {Wilkins}}]{Trussler2025}
{Trussler}, J. A.~A., {Conselice}, C.~J., {Adams}, N., {et~al.} 2025, \mnras, 537, 3662

\bibitem[{{Valentino} {et~al.}(2023){Valentino}, {Brammer}, {Gould}, {Kokorev}, {Fujimoto}, {Jespersen}, {Vijayan}, {Weaver}, {Ito}, {Tanaka}, {Ilbert}, {Magdis}, {Whitaker}, {Faisst}, {Gallazzi}, {Gillman}, {Gim{\'e}nez-Arteaga}, {G{\'o}mez-Guijarro}, {Kubo}, {Heintz}, {Hirschmann}, {Oesch}, {Onodera}, {Rizzo}, {Lee}, {Strait}, \& {Toft}}]{Valentino2023}
{Valentino}, F., {Brammer}, G., {Gould}, K. M.~L., {et~al.} 2023, \apj, 947, 20

\bibitem[{{Valentino} {et~al.}(2025){Valentino}, {Heintz}, {Brammer}, {Ito}, {Kokorev}, {Whitaker}, {Gallazzi}, {de Graaff}, {Weibel}, {Frye}, {Kamieneski}, {Jin}, {Ceverino}, {Faisst}, {Farcy}, {Fujimoto}, {Gillman}, {Gottumukkala}, {Hamadouche}, {Harrington}, {Hirschmann}, {Jespersen}, {Kakimoto}, {Kubo}, {Lagos}, {Lee}, {Magdis}, {Man}, {Onodera}, {Rizzo}, {Shimakawa}, {Setton}, {Tanaka}, {Toft}, {Wu}, \& {Zhu}}]{Valentino2025}
{Valentino}, F., {Heintz}, K.~E., {Brammer}, G., {et~al.} 2025, \aap, 699, A358

\bibitem[{{Valentino} {et~al.}(2020){Valentino}, {Tanaka}, {Davidzon}, {Toft}, {G{\'o}mez-Guijarro}, {Stockmann}, {Onodera}, {Brammer}, {Ceverino}, {Faisst}, {Gallazzi}, {Hayward}, {Ilbert}, {Kubo}, {Magdis}, {Selsing}, {Shimakawa}, {Sparre}, {Steinhardt}, {Yabe}, \& {Zabl}}]{Valentino2020}
{Valentino}, F., {Tanaka}, M., {Davidzon}, I., {et~al.} 2020, \apj, 889, 93

\bibitem[{{van Dokkum} {et~al.}(2014){van Dokkum}, {Bezanson}, {van der Wel}, {Nelson}, {Momcheva}, {Skelton}, {Whitaker}, {Brammer}, {Conroy}, {F{\"o}rster Schreiber}, {Fumagalli}, {Kriek}, {Labb{\'e}}, {Leja}, {Marchesini}, {Muzzin}, {Oesch}, \& {Wuyts}}]{vanDokkum2014}
{van Dokkum}, P.~G., {Bezanson}, R., {van der Wel}, A., {et~al.} 2014, \apj, 791, 45

\bibitem[{{Vogelsberger} {et~al.}(2020){Vogelsberger}, {Marinacci}, {Torrey}, \& {Puchwein}}]{Vogelsberger2020}
{Vogelsberger}, M., {Marinacci}, F., {Torrey}, P., \& {Puchwein}, E. 2020, Nature Reviews Physics, 2, 42

\bibitem[{{Weibel} {et~al.}(2025){Weibel}, {de Graaff}, {Setton}, {Miller}, {Oesch}, {Brammer}, {Lagos}, {Whitaker}, {Williams}, {Baggen}, {Bezanson}, {Boogaard}, {Cleri}, {Greene}, {Hirschmann}, {Hviding}, {Kuruvanthodi}, {Labb{\'e}}, {Leja}, {Maseda}, {Matthee}, {McConachie}, {Naidu}, {Roberts-Borsani}, {Schaerer}, {Suess}, {Valentino}, {van Dokkum}, \& {Wang}}]{Weibel2024qgal}
{Weibel}, A., {de Graaff}, A., {Setton}, D.~J., {et~al.} 2025, \apj, 983, 11

\bibitem[{{White} \& {Rees}(1978)}]{White1978}
{White}, S.~D.~M. \& {Rees}, M.~J. 1978, \mnras, 183, 341

\bibitem[{{Wilkins} {et~al.}(2011){Wilkins}, {Bunker}, {Stanway}, {Lorenzoni}, \& {Caruana}}]{Wilkins2011}
{Wilkins}, S.~M., {Bunker}, A.~J., {Stanway}, E., {Lorenzoni}, S., \& {Caruana}, J. 2011, \mnras, 417, 717

\bibitem[{{Wilkins} {et~al.}(2019){Wilkins}, {Lovell}, \& {Stanway}}]{Wilkins2019}
{Wilkins}, S.~M., {Lovell}, C.~C., \& {Stanway}, E.~R. 2019, \mnras, 490, 5359

\bibitem[{{Witten} {et~al.}(2025{\natexlab{a}}){Witten}, {McClymont}, {Laporte}, {Roberts-Borsani}, {Sijacki}, {Tacchella}, {Simmonds}, {Katz}, {Ellis}, {Witstok}, {Maiolino}, {Ji}, {Hayes}, {Looser}, \& {D'Eugenio}}]{Witten2025}
{Witten}, C., {McClymont}, W., {Laporte}, N., {et~al.} 2025{\natexlab{a}}, \mnras, 537, 112

\bibitem[{{Witten} {et~al.}(2025{\natexlab{b}}){Witten}, {Oesch}, {Bennett}, {Meyer}, {Giovinazzo}, {Covelo-Paz}, {Baker}, \& {Ivey}}]{Witten2025c}
{Witten}, C., {Oesch}, P.~A., {Bennett}, J.~S., {et~al.} 2025{\natexlab{b}}, arXiv e-prints, arXiv:2511.05647

\bibitem[{{Witten} {et~al.}(2025{\natexlab{c}}){Witten}, {Oesch}, {McClymont}, {Meyer}, {Fudamoto}, {Sijacki}, {Laporte}, {Bennett}, {Simmonds}, {Giovinazzo}, {Danhaive}, {Ciesla}, {Carvajal-Bohorquez}, \& {Trebitsch}}]{Witten2025b}
{Witten}, C., {Oesch}, P.~A., {McClymont}, W., {et~al.} 2025{\natexlab{c}}, arXiv e-prints, arXiv:2507.06284

\bibitem[{{Wu}(2025)}]{Wu2025}
{Wu}, P.-F. 2025, \apj, 978, 131

\bibitem[{{Zavala} {et~al.}(2018){Zavala}, {Monta{\~n}a}, {Hughes}, {Yun}, {Ivison}, {Valiante}, {Wilner}, {Spilker}, {Aretxaga}, {Eales}, {Avila-Reese}, {Ch{\'a}vez}, {Cooray}, {Dannerbauer}, {Dunlop}, {Dunne}, {G{\'o}mez-Ruiz}, {Micha{\l}owski}, {Narayanan}, {Nayyeri}, {Oteo}, {Rosa Gonz{\'a}lez}, {S{\'a}nchez-Arg{\"u}elles}, {Schloerb}, {Serjeant}, {Smith}, {Terlevich}, {Vega}, {Villalba}, {van der Werf}, {Wilson}, \& {Zeballos}}]{Zavala2018}
{Zavala}, J.~A., {Monta{\~n}a}, A., {Hughes}, D.~H., {et~al.} 2018, Nature Astronomy, 2, 56

\bibitem[{{Zhang} {et~al.}(2025){Zhang}, {de Graaff}, {Setton}, {Price}, {Bezanson}, {Lagos}, {Cutler}, {McConachie}, {Cleri}, {Cooper}, {Gottumukkala}, {Greene}, {Hirschmann}, {Khullar}, {Labbe}, {Leja}, {Maseda}, {Matthee}, {Miller}, {Nanayakkara}, {Suess}, {Wang}, {Whitaker}, \& {Williams}}]{Zhang2025}
{Zhang}, Y., {de Graaff}, A., {Setton}, D.~J., {et~al.} 2025, arXiv e-prints, arXiv:2508.08577

\bibitem[{{Zinger} {et~al.}(2020){Zinger}, {Pillepich}, {Nelson}, {Weinberger}, {Pakmor}, {Springel}, {Hernquist}, {Marinacci}, \& {Vogelsberger}}]{Zinger2020}
{Zinger}, E., {Pillepich}, A., {Nelson}, D., {et~al.} 2020, \mnras, 499, 768

\end{thebibliography}

\begin{appendix}

\end{appendix}

\end{document}